\documentclass[aps,twocolumn,showpacs,superscriptaddress,citeautoscript,prb,floatfix]{revtex4-2}

\usepackage{graphicx} % Required for inserting images
\usepackage{amsmath}
\usepackage{amsfonts}
\usepackage{tabularx}
\usepackage{cancel}
\usepackage{bbold} % para la identidad
\usepackage{wrapfig}

\usepackage{amssymb}
\usepackage{bm}
\usepackage[T1]{fontenc}
\usepackage{textcomp}
\usepackage{enumitem}
\usepackage{cancel}
\usepackage{bbold}
\usepackage{esvect}
\usepackage{commath}
\usepackage[most]{tcolorbox}
\usepackage{lipsum, babel}
\usepackage{comment}
\usepackage{lineno}

\usepackage{xcolor}

\usepackage{adjustbox}

\newcommand{\angstr}{\textup{~\AA}}

\usepackage[left=23mm,right=23mm, top = 20mm, bottom = 20mm, paper = a4paper]{geometry}

\begin{document}

%\title{Topology and disorder in a 2D semi-Dirac material}
\title{High order momentum topological insulator in 2D semi-Dirac materials}
\author{Marta Garc\'{\i}a Olmos}
\email{mgarcia.o@usal.es}

\affiliation{%
 Nanotechnology Group, USAL—Nanolab, University of Salamanca\\
 Plaza de la Merced, Edificio Triling\"{u}e, 37008, Salamanca, Spain.
}%

\affiliation{Instituto de Estructura de la Materia IEM-CSIC, Serrano 123, E-28006 Madrid, Spain}

\author{Yuriko Baba}
\email{yuribaba@ucm.es}

\affiliation{
Department of Theoretical Condensed Matter Physics, Condensed
Matter Physics Center (IFIMAC)
Universidad Autónoma de Madrid, E-28049 Madrid, Spain}

\affiliation{GISC, Departamento de F\'{\i}sica de Materiales, Universidad Complutense, E-28040 Madrid, Spain}

\author{Mario Amado}
\email{mario.amado@usal.es}
\affiliation{%
 Nanotechnology Group, USAL—Nanolab, University of Salamanca\\
 Plaza de la Merced, Edificio Triling\"{u}e, 37008, Salamanca, Spain.
}%

\author{Rafael A. Molina}
\email{rafael.molina@csic.es}
\affiliation{Instituto de Estructura de la Materia IEM-CSIC, Serrano 123, E-28006 Madrid, Spain}

%\date{July 2023}
\
\begin{abstract}
Semi-Dirac materials in 2D present an anisotropic dispersion relation, linear along one direction and quadratic along the perpendicular one. This study explores the topological properties and the influence of disorder in a 2D semi-Dirac Hamiltonian. Anisotropic edge states appear only in one direction. Their topological protection can be rigorously founded on the Zak phase of the one-dimensional reduction of the semi-Dirac Hamiltonian, parametrically depending on one of the momenta. In general, only a single value of the momentum is topologically protected so these systems can be considered as high order momentum topological insulators. We explore the dependence on the disorder of the edge states and the robustness of the topological protection in these materials. We also explore the consequences of the high order topological protection in momentum space for the transport properties in a two-terminal configuration.  

\end{abstract}

\maketitle

\section{Introduction}

%**INTERESANTE: \url{https://arxiv.org/pdf/1507.01742.pdf}
% include the berry curvature plot.

%The condensed matter field has been revolutionized with the emergence of two-dimensional materials and heterostructures, such as Dirac and semi-Dirac materials. These materials exhibit unconventional electronic behavior owing to their unique band structures, leading to protected surface states and exciting quantum phenomena that deviate from the behavior of the materials with the usual quadratic dispersion. Some examples are the the family of quantum Hall effects \cite{QHEreview, QSHEreview}\\

%Dirac materials present linear dispersion in their bands so relativistic carriers that give rise to quantized response and possible protected surface states due to the non-trivial topology of the bulk states. The first and paradigmatic example of Dirac material in 2D is graphene which is a one-atom-thick carbon sheet on a honeycomb lattice \cite{Novoselov2004,CastroNeto2009}. Other interesting Dirac materials, the topological insulators, discovered soon after, are insulators in the bulk but conductors at their surface due to the non-trivial topological structure of the bands around the Fermi energy \cite{Koenig2007,Qi2011}. 

Condensed Matter physics has been revolutionized with the emergence of two-dimensional materials and heterostructures. These materials often exhibit unconventional behavior due to their unique band structures that deviate from the usual quadratic dispersion becoming linear around the Fermi energy. The charge carriers behave then as relativistic massless fermions and can give rise to quantized transport properties and possible protected surface states due to non-trivial topology of the bulk electronic band structure. The first and paradigmatic example of such Dirac material in 2D is graphene which is a one-atom-thick carbon sheet on a honeycomb lattice \cite{Novoselov2004,CastroNeto2009}. Since then, the exploration of novel materials has expanded, revealing several examples with intringuing topological features such as topological insulators \cite{Koenig2007,Qi2011,QSHEreview} and Weyl semimetals \cite{Burkov2011,Liu_2014, Naumann2021}.
% Other interesting Dirac materials, the topological insulators, discovered soon after, are insulators in the bulk but conductors at their surface \cite{Koenig2007,Qi2011,QSHEreview}.

% https://arxiv.org/pdf/2303.04468.pdf
In contrast, semi-Dirac 2D materials display a linear dispersion along one direction and a quadratic one along the perpendicular direction of the momentum space. This  dispersion has been predicted in a wide variety of materials, VO$_2$/TiO$_2$ nanostructures \cite{Pardo2009}, $\alpha$-(BEDT-TTF)$_2$I$_3$ salts under pressure \cite{Katayama2006}, phosphorene \cite{CastellanosGomez2014, black_ph_strain}, thin films of Cd$_3$As$_2$ \cite{Liu2022} and silicene oxide \cite{Xiao_shotnoise}, among others.
% https://sci-hub.et-fine.com/10.1063/1.3560505 Ti$_3$SiC$_2$ embedded in SiC
Uniaxial strain applied on graphene, or, in general, a monoatomic honeycomb lattice induces also a semi-Dirac dispersion \cite{Montambaux2009b}. Several theoretical studies have addressed the different transport, optical and magnetic properties of these types of materials
\cite{Banerjee2009, Montambaux2009, Carbotte2019, Huang2023TheGA}. 
%**https://www.semanticscholar.org/reader/f8518918d3c0fef6752dc1443af11387d28fa04f

The potential for these materials to exhibit topological properties lies in the presence or absence of band inversion \cite{Qi2011,Zhu2012,Shen2017_Book}. In 2D crystalline materials the standard topological number for evaluating topological properties is the Chern number that can be computed by integrating the Berry curvature in the first Brillouin zone. A finite Chern number $C$ implies a quantized non-zero anomalous Hall conductance proportional to $C$ \cite{Thouless1982, Liu2016}. However, due to their anisotropic nature, two-dimensional topological invariants cannot fully capture the topological properties of semi-Dirac materials and the Chern number for these materials is zero. 
\cite{Montambaux2009, Montambaux2009b, Huang2015}. The anisotropic band structure in the standard semi-Dirac model comes from the merging of two Dirac cones \cite{Montambaux2009}. A new type-II semi-Dirac dispersion coming from the merging of three Dirac points has been proposed and applied to VO$_2$/TiO$_2$ nanostructures. For that case, in the presence of spin-orbit coupling the Chern number obtained is $C=-2$ \cite{Huang2015}. In our work, we do not consider such model and concentrate on the type-I semi-Dirac model that can be derived from a strained honeycomb lattice, see for example \cite{GuineaSols2008, Delplace_2011}. 
% https://iopscience.iop.org/article/10.1088/1367-2630/10/10/103027/pdf
% https://www.researchgate.net/publication/340500438_Semi-Dirac_transport_and_anisotropic_localization_in_polariton_honeycomb_lattices
% https://www.nature.com/articles/srep17571

Our idea is then to reinterpret the problem in terms of one-dimensional invariants parametically depending on the momentum of one specific spatial direction. Thus, it is possible to compute the Zak phase \cite{ZakPRL,Shen2017_Book}, a standard one-dimensional topological number, as a function of one of the momenta of the system to study the topological properties of our model. We find that the Zak phase is non-trivial only for the zero momentum state where the eigenstates are degenerate. This is consistent with the fact that the dispersion of the edge states appearing for the type-I semi-Dirac model is quadratic and not linear and with the Chern number being zero. Although, edge states appear in one of the directions for a finite semi-Dirac nanoribbon only one single state is topologically protected. This is comparable to the behavior of high order topological insulators \cite{Schindler2018} where the topological protection is based on a dimensional reduction of the Hamiltonian \cite{Trifunovic2019}. However, for the model we consider this occurs in momentum space instead of the spatial dimensions. 

%This fact implies than in perfectly clean samples one should find quantized anomalous quantum Hall conductance in semi-Dirac materials but the quantization should be easily destroyed by disorder. 
This fact implies that, although, we do not have anomalous quantum Hall conductance in semi-Dirac materials, we should observe an anisotropic robust quantization of the two-terminal conductance for small bias due to the topologically protected zero mode state. We have explored the interplay of topology and disorder in semi-Dirac materials by studying the properties of our model as a function of disorder, including the participation ratio of the edge states and the two-terminal conductance. We find that the states closer to zero momentum in one of the directions are more robust to disorder perturbations in consistency with the results from the topological analysis.

The paper is organized as follows. In Sec. II, we introduce the model and an analytical approach to characterize the edge states. In Sec. III we clarify the definition of the topological invariant for semi-Dirac materials and we prove that strict topological protection only applies to a zero momentum state in one direction. Then, we analyze the resilience of edge states to disorder. In Sec. IV we introduce two different models of disorder and study its effect on the expected value of the position in the direction perpendicular to the decay of the edge states and the conductance.

%%%%%% Conforme avancemos vemos cómo continúan las secciones.

\section{Semi-Dirac model}

Our model represents a two-banded type-I semi-Dirac material in two-dimensions
where we have included a mass term. To represent this system we consider a $k \cdot p$ Hamiltonian, generically expressed for two bands as $H_0(\boldsymbol{k}) = \vec{d} (\boldsymbol{k}) \cdot \vec{\sigma}$. This Hamiltonian encapsulates the fundamental low-energy physics of our system. We do not include terms proportional to the unit matrix as they do not impact the topological properties we are interested in. Here, $\vec{d} (\boldsymbol{k})$ is a vector in the momentum space, and $\vec{\sigma}$ is a vector containing the Pauli matrices. In the semi-Dirac case, the components of the vector $\vec{d}(\boldsymbol{k})$ are written in the following form 
\begin{equation} 
    \vec{d} (\boldsymbol{k}) = \begin{pmatrix}
        V_x k_x^2 \\
        V_y k_y\\
        M_0 - M_{1x} k_x^2 - 
 M_{1y} k_y^2
    \end{pmatrix} ,
    \label{eq:ham_SD}
\end{equation}
where all the parameters are real. The eigenvalues and eigenvectors can be expressed in terms of the $\boldsymbol{d} (\boldsymbol{k})$-vector generically as 
\begin{align}
    E_{\pm} &= \pm d
    % |\boldsymbol{d}(\boldsymbol{k})|,\\
    \Psi^{\pm} &= \frac{1}{\sqrt{2d(d\pm d_z)}}\begin{pmatrix} d_z \pm d \\ d_x + i d_y \end{pmatrix} ,
\end{align}
with $d=|\boldsymbol{d}(\boldsymbol{k})|$.

\begin{figure*}[t]
    \raggedright
    \includegraphics[width=0.9\textwidth]{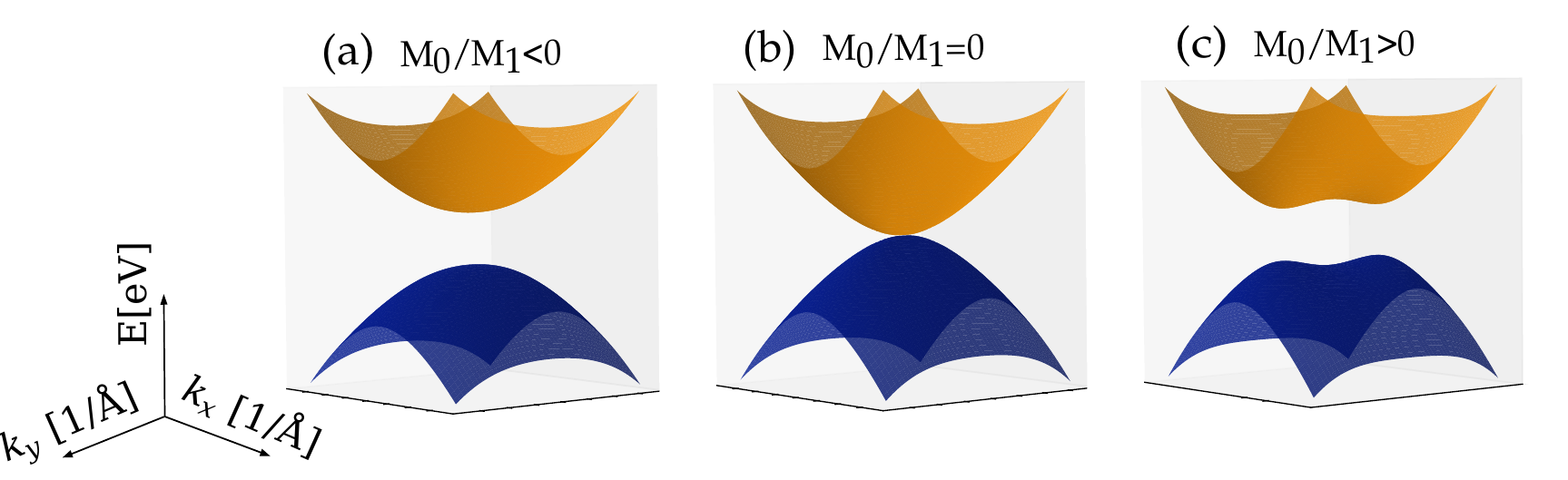}
     \caption{Band structure of the bulk states associated to the semi-Dirac Hamiltonian \eqref{eq:ham_SD} in terms of the $\text{sgn} (M_0/M_1)$. For $M_0 = 0$ (b), the gap closes, distinguishing the transition between the trivial regime, (negative sign of $M_0/M_1$, (a)) and the band-inversion regime (positive sign of $M_0/M_1$, (c)).}
%%%%% Incluiría M_0/M_1 >0 =0 <0 en la propia figura
     \label{fig:bulkbands}
\end{figure*}
% \begin{figure}
% \centering
% \begin{subfigure}{0.3\textwidth}
%     \includegraphics[width=0.5\columnwidth]{bulk_gapbands_cropped.pdf}
%     \caption{Band structure of the bulk states associated to the Semi-dirac model \eqref{eq:ham_SD}.}
%     \label{fig:bulkbands}
% \end{subfigure}
% \hfill
% \begin{subfigure}{0.3\textwidth}
%     \includegraphics[width=0.5\columnwidth]{bulk_bands_Vxzero_cropped.pdf}
% \end{subfigure}

The minimal model for semi-Dirac Hamiltonians does not includes the cuadratic terms with $V_x$ and $M_{1y}$. However, taking into account symmetry considerations, these terms should be possible for most materials. 
%Although the term with $V_x$ is not very important for what follows, 
The term with $M_{1y}$ is essential when considering band inversion and the topological properties while the term with $V_x$ controls the dispersion properties close to $k_x=0$.   
The bulk band structure of the model for a case with $M_{1x}=M_{1y}=M_1$ is displayed schematically in Fig. \ref{fig:bulkbands}. The spectrum is gapped due to the mass term $M_0$. The change in the sign of the $M_0/M_1$ ratio closes and reopens the gap, switching the bands from a regular gap to an inverted one as shown in the different panels of the figure. The dispersion relation exhibits a quadratic dependence in the $x$-direction of the momentum space and a linear dependence in the $y$-direction. This anisotropic dispersion reflects the different behaviors of charge carriers along the two directions in momentum space. 
%However, the presence of the off-diagonal term $V_x$ introduce coupling between different states leading to a non-linear dependence along the $k_y$ direction of the bands. 

\subsection{Characterization of the edge states}

% If you make $V_x = 0$ or impose a linear dependence in $k_x$ the Berry curvature of one Dirac cone is monopole-like, so because of translational symmetry the system will be forced to have two Dirac points (**with different chirality).\\

% Hay un control de versiones descargado, modifico esta nueva versión con las anotaciones de Yuriko.

In this section we derive analytically the edge states for a nanoribbon with finite width in $y$-direction and periodic boundary conditions in $x$-direction.
For simplicity, we consider each edge separately, treating them as isolated edges of the infinite half-planes defined by $y \geq W/2$ for the upper edge and $y \leq -W/2$ for the lower edge.
% For simplicity, we are going to study the two edges separately, this means that we are imposing Dirichlet boundary conditions at the two edges $y = +  W / 2 $  and $y = -  W / 2$ separately, considering that the system is semi-infinite for $y<W/2$ and $y>-W/2$, respectively. 
This approximation suites the regime of a nanoribbon with width $W \gg \lambda$ being $\lambda$ the decay length of the states.
In this regime, both edges can be solved separately neglecting the hybridization generated by the finite size in $y$-direction.\\

We explore solutions that decay exponentially from the edge in the confined direction and travel as a plane-wave along the $x$-direction, \text{$\psi^{\pm}_{k_x}$ \footnotesize $\sim e^{ik_x x} e^{ \mp \lambda ( \pm W/2 - y)} (\alpha~,~\beta)^T$} where the decay length satisfies $\text{Re}(\lambda) > 0$. The two signs refers to the two edges located at $y= \pm W/ 2$ respectively, $\alpha$, $\beta$ and $\lambda$ are complex numbers. \\

Solving the eigenvalue problem yields to the same expression for both edges,
%
%\begin{equation}
%E = \pm \sqrt{\left[M_0 - M_1(k_x^2 - \lambda^2)\right]^2 + V_x^2 k_x^4 - V_y^2 \lambda^2} .
%\end{equation}
%
\begin{equation}
E = \pm \sqrt{\left[M_0 - M_{1x}k_x^2 - M_{1y}\lambda^2)\right]^2 + V_x^2 k_x^4 - V_y^2 \lambda^2} .
\end{equation}
which results in four roots, $\pm \lambda_1$ and $\pm \lambda_2$ whose explicit expressions are included in Appendix \ref{ap:edge-states}. Note that the $\lambda_i$ for $i=1,2$ are functions of the model parameters and they depend explicitly on the momenta $k_x$. 

To ensure the requiered exponential decay positive values of $\lambda$ describe edge states decaying from the upper edge while the negative ones describe edge states decaying from the lower edge. The solution then takes the general form,
\begin{equation}
    \psi^{\pm}_{k_x} = A_1 \begin{pmatrix}
        \alpha_1 \\ \beta_1
    \end{pmatrix} e^{\mp \lambda_1 (\pm W/2 - y)} + A_2 \begin{pmatrix}
        \alpha_2 \\ \beta_2
    \end{pmatrix} e^{\mp \lambda_2 (\pm W/2 - y)} ,
\end{equation}
where $\alpha_i = V_x k_x^2 \mp V_y \lambda_i$ and $\beta_i = E - M_0 + M_{1x}k_x^2 - M_{1y}\lambda_i^2$ and $A_i$ are normalization constants.\\

By employing the proposed \textit{ansatz} in the Schr\"odinger equation with Dirichlet boundary conditions $\psi_{k_x}^{\pm} (y=\pm W/2) = 0$, we obtain the dispersion relation for the edge states as a function of the decay lengths \eqref{eq:decay_length_i},
\begin{equation}\label{eq:edgedispersion}
   % E^{\pm} = M_0 - M_1k_x^2 - M_1(\lambda_1 \lambda_2) \mp M_1\frac{V_x}{V_y}k_x (\lambda_1 + \lambda_2). 
    E^{\pm} = M_0 - M_{1x}k_x^2 - M_{1y}(\lambda_1 \lambda_2) \mp M_{1y}\frac{V_x}{V_y}k_x (\lambda_1 + \lambda_2). 
    %\\ \nonumber &\simeq \left. 
    %E \right|_{k_x=0} + \left. \frac{d^2E^-}{dk_x^2}  \right|_{k_x = 0} k_x^2 + O(k_x^4). 
\end{equation}
In order to clarify the dependency on the momentum $k_x$, it is illustrative to perform an expansion around $k_x = 0$. This way we obtain a quadratic dispersion in $k_x$ 
%And by performing an expansion around $k_x = 0$, we check analytically that the spectrum depends quadratically on $k_x$,
\begin{equation} \label{eq:dispers_relat}
    E^{\pm} \simeq M_0 +  M_{1y}  
    \left( \lambda_1^{(0)} \lambda_2^{(0)} \right) ~\mp~ M_{1y} \frac{V_x}{V_y} 
    \left(\lambda_1^{(0)} + \lambda_2^{(0)} \right)k_x^2 .
\end{equation} 
Here, $\lambda_i^{(0)}$ are two constants that depend only on the Hamiltonian parameters obtained by evaluating the decay lengths at $k_x = 0$.\\

There are two things that we want to highlight about these results.
Firstly, non-lineal dispersion and the complicated dependence of the spinor coefficients imply that the edge states are not chiral which is consistent with the value of $C=0$ of the Chern number.
Secondly, in this two-band Hamiltonian there is a correspondence between the sign of the decay length and the sign of the energy, which are opposite.
In other words, the wavefunction $\psi^-$ corresponds to edge states decaying in the positive half-plane, i.e. for $y> - W/2$, and is associated with negative energy values. The opposite is true for negative values of $\lambda$, so the electronic or hole character of the edge states depend on whether they are in the upper or lower edge of the ribbon.
%\marta{Signos revisados. Como había un poco de lío entre semi-planos y><0 lo he cambiado todo a semi-planos y > -W/2, y < W/2 consistentemente con el sistema numérico, no sé si queda más lioso.} \yuriko{Está genial!}
%Second, in this two-band Hamiltonian positive values of $\lambda$ corresponding to edge states decaying to $y<0$ correspond to positive values of the energy while the opposite is true for negative values of $\lambda$ so the electronic or hole character of the edge states depend on whether they are in the upper or lower edge of the ribbon.
The lack of chirality indicate a symmetry imbalance, in order to restore the symmetry, a $4 \times 4$ Hamiltonian with opposite chirality would be needed. Thus, spin degrees of freedom are incorporated, allowing the inclusion of Kramer pairs similarly to the case of the quantum spin Hall models \cite{QSHEreview}. 
% *** What symmetry are we talking about? TRS??
% In order to restore the symmetry, a $4 \times 4$ Hamiltonian with opposite chirality is needed similarly as in the case of the spin Hall effect models \cite{QSHEreview}. 

\begin{figure}[h]
  \begin{minipage}{\columnwidth}
    \centering
    \includegraphics[width=\columnwidth]{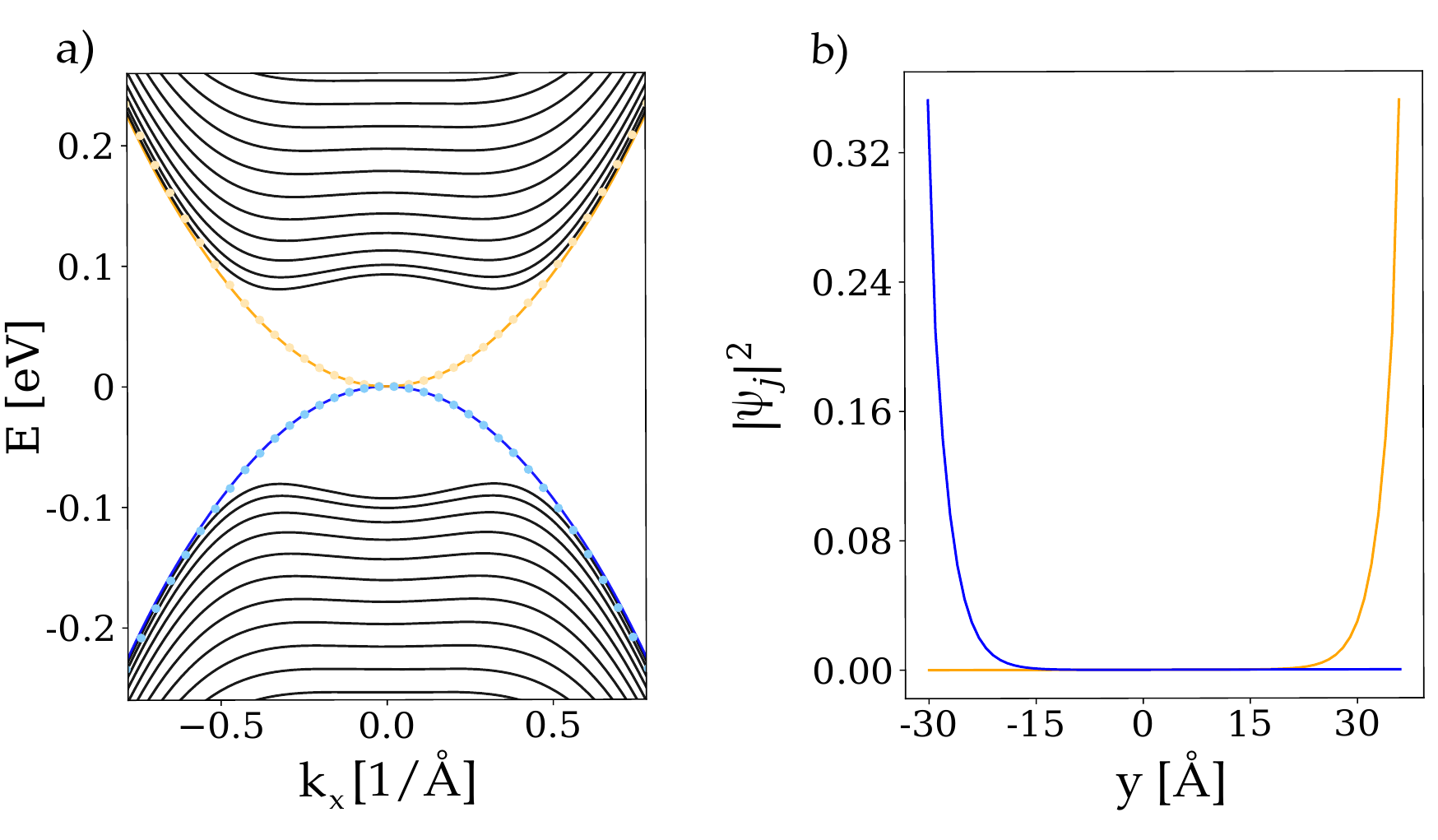}
    \caption{Panel (a): band structure of a finite size system with translational symmetry in $x$-direction. The solid line corresponds to the numerical result and the dotted line is the analytical result for the edge state.
    Panel (b): probability density of the edge states for $k_x=8\cdot 10^{-3}~\text{\AA}^{-1}$ in terms of the sites of the nanoribbon.
    The orange line corresponds to the edge state with positive energy, $E = 5 \cdot 10^{-4}~\mathrm{eV}$ and localization in the upper edge. Conversely, the state with negative energy $E = - 5 \cdot 10^{-4}~\mathrm{eV}$ is depicted in in blue and it is localized in the lower edge.}
    % it is k_x = 0.037
    \label{fig:bands_finitesyst_ydir}
  \end{minipage}
\end{figure}

\begin{figure}[h]
  \begin{minipage}{\columnwidth}
    \centering
    \includegraphics[width=0.55\columnwidth]{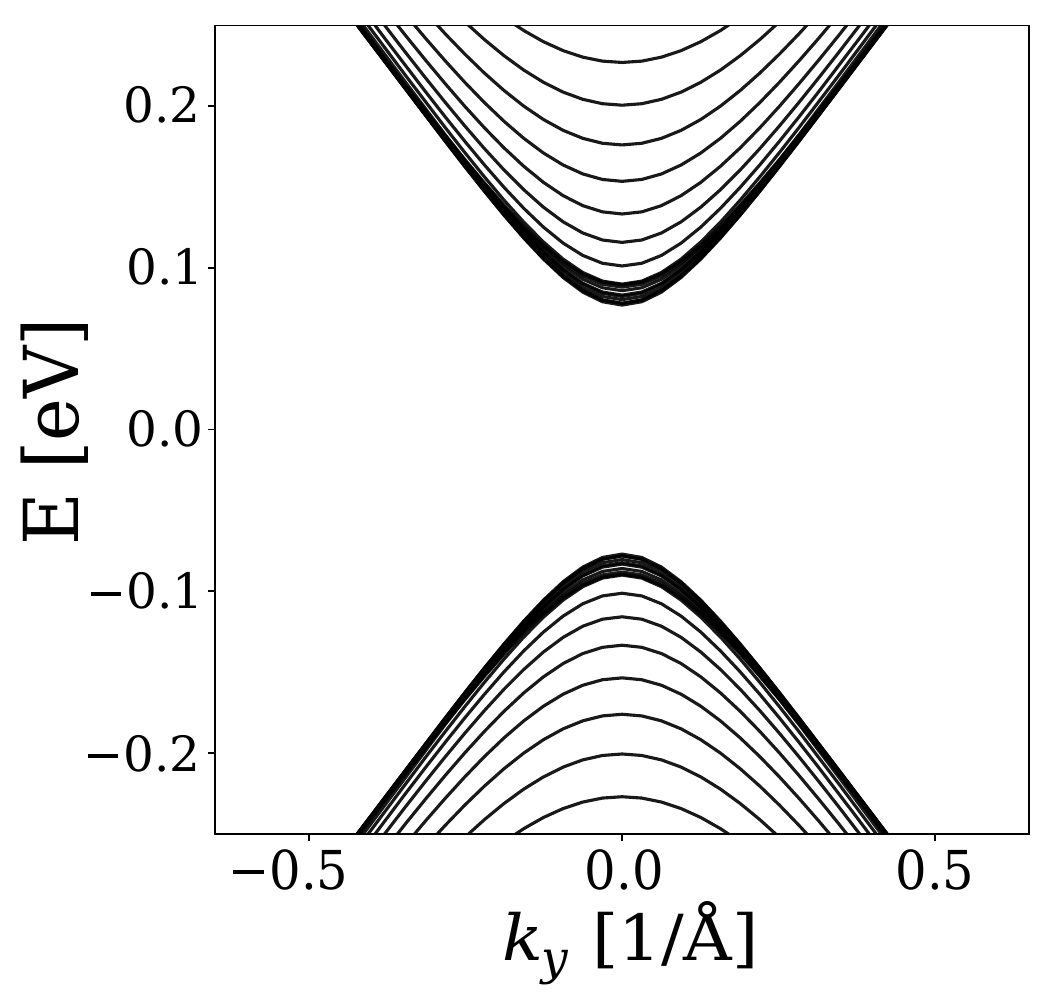}
    \caption{Bands of a system with translational symmetry in $y$-direction and boundaries in $x$-direction.}
    \label{fig:bands_xdir}
  \end{minipage}
\end{figure}

To further investigate the nature of these states, we have relied on numerical simulations.
The simulations are based on nearest-neighbor tight-binding discretization of the model described by Eq. \ref{eq:ham_SD} in a square lattice.
To implement this, we use Kwant library \cite{Groth_2014}, an open-source python-package that efficiently can be employed also for the transport calculations exposed in the following section.
%by making the substitutions $k_i \rightarrow \frac{1}{a} \sin (k_i a)$, $k_i^2 \rightarrow \frac{2}{a^2} (1 - \cos(k_i a))$ \cite{Shen2017_Book}. 
All figures have been made with the set of parameters included in table \ref{tab:real_complex_params}, for which there is band inversion.
\begin{table}[h] 
    \begin{center}
        \begin{tabular}{| c | c | c | c | c |} 
        \hline 
        $M_0 \left[ \text{eV} \right]$  & $M_{1x} [ \text{eV}/\text{ \AA}^2 ]$ &$M_{1y} [ \text{eV}/\text{ \AA}^2 ]$ &$V_x [ \text{eV}/\text{ \AA}^2 ]$ & $V_y [ \text{eV}/\text{ \AA} ] (\mathbb{R} / \mathbb{C})$ \\ \hline
        0.09 & 0.23 & 0.23 & -0.38 & -0.5 / -0.18 \\ \hline
        \end{tabular}
        \caption{Set of parameters of the model in the band inversion regime used for the figures. The two values of $V_y$ correspond to a case with exponential decay and a case with oscillatory decay. See the text for details. We have used $V_y=-0.5~\mathrm{eV /\angstr} $  except in the right panel of Fig. \ref{fig:gap_width}.}
        \label{tab:real_complex_params}
    \end{center}
\end{table}
The left panel of Fig. \ref{fig:bands_finitesyst_ydir} shows the bands of a nanoribbon of width $W=66$ \AA ~ centered at $y = 0$ and with translational symmetry in the $x$-direction, so $k_x$ is a good quantum number. In the right panel we show the probability density of the edge states for 
$k_x=0.04~\text{\AA}^{-1}$, the upper edge corresponds to electronic excitations and the lower one to hole excitations.
In this semi-infinite system, the edge states are degenerate for $k_x=0$ since the term that couples them, $V_x k_x^2$, disappear. The width of the gap due to finite size effects is so small for the width considered that it cannot be seen in the figure. We discuss this gap in the following paragraph. The dots in the figure correspond to the analytical expression for the dispersion of the edge states from Eq. \ref{eq:edgedispersion}.
The agreement between the analytical results and the numerical simulations is excellent as the nanoribbon width is large enough to show uncoupled edges.
Additionally, in Fig. \ref{fig:bands_xdir} we present the spectrum associated with the confined states in the $x$-direction with the same width and with periodic boundary conditions in the $y$-direction, to show that there are no edge states due to the quadratic dependence in this direction.
\begin{figure}[h]
      \includegraphics[width=\linewidth]{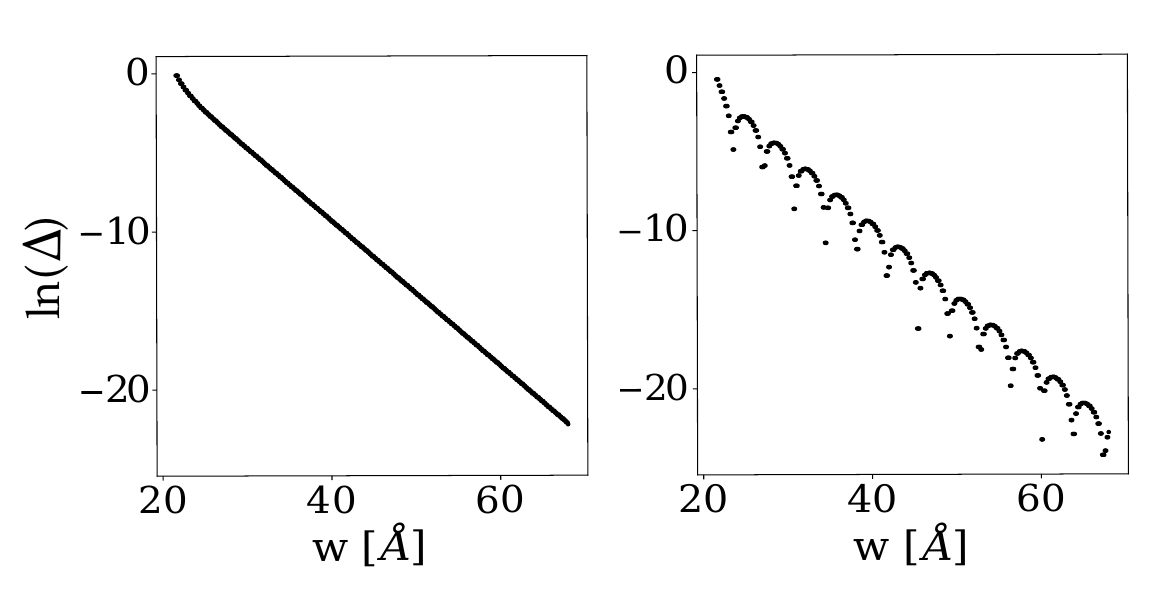}
    \caption{Natural logarithm of the gap versus the width of the system in a nanoribbon of width $W$ and periodic boundary conditions in the $x$-direction. While in the real case the gap decays exponentially, in the complex case it decays oscillatorily. The parameters used in the simulations are collected in Table \ref{tab:real_complex_params} with $V_y=-0.5~\mathrm{eV/}\angstr$ in the left panel and $V_y=-0.18~\mathrm{eV/}\angstr$ in the right panel.}
    \label{fig:gap_width}
\end{figure}

From the previous results, we can differentiate two cases depending on the nature of the decay length of the states, if $V_y^2<4M_0M_{1y}$ the decay lengths have a non-zero complex part, while for $V_y^2>4M_0M_{1y}$ , they are purely real.
This situation is similar in other topologically protected edge states \cite{Gonzalez2017}.

For finite systems the real or complex nature of $\lambda$ is going to determine how the width of the gap between edge states at $k_x=0$ due to finite size effects changes with the width of the system \cite{Benito2019}.
For $\lambda$ purely real the gap has an exponential dependence with the width of the system while for complex $\lambda$ the dependence is oscillatory. In Fig. \ref{fig:gap_width} we compare the behavior of the finite size gap between the complex and the purely real values of $\lambda$ for the two values of $V_y$ included in table \ref{tab:real_complex_params}.

\section{Topological character}

%**Our system has time-reversal symmetry but lacks (spatial-)inversion symmetry 
%because the term of $\sigma_z$ is even so, by symmetry considerations, it can be a candidate to have non-trivial topology \cite{Berry_phase_review}. 
% The components of the Berry curvature for the two band Hamiltonian \eqref{eq:ham_SD} in terms of the $d$-vector can be express as follows
% \begin{equation}
%     \Omega_{ij} (\boldsymbol{k}) = \frac{1}{2} \frac{\boldsymbol{d} (\boldsymbol{k}) \cdot (\partial_i \boldsymbol{d} (\boldsymbol{k}) \times \partial_j \boldsymbol{d} (\boldsymbol{k}))}{d^3(\boldsymbol{k})} .
%     \label{eq:Berry}
% \end{equation}
% The topological character of the states is related with the phase acquired during an adiabatic evolution across the Brillouin zone. 

In this section, we investigate the particular topological protection of the edge states appearing in our system.
As previously mentioned, due to the anisotropic nature of the semi-Dirac model, the two-dimensional invariants do not capture the topological properties. Instead, the relevant topological aspects are associated to the geometric phase acquiered during the adiabatic evolution of a quantum state along a closed path in the $k_y$-direction, rather than in the entire Brillouin zone. This is directly linked to the Zak phase, a one-dimensional invariant that becomes a key element revealing the system's topological intricacies.
For the $n-$band, it is defined as \cite{ZakPRL},
\begin{equation} \label{eq: berry connection, d-form}
    \mathcal{Z}^n = i \oint dk \left\langle u_n(\boldsymbol{k}) \right| \partial_{k} \left| u_n (\boldsymbol{k}) \right\rangle ,
\end{equation}
% \begin{equation}
%     \mathcal{Z}^{\pm} = i \oint dk \frac{i (d_x \partial_i d_y - d_y \partial_i d_x)}{2d(d \pm d_z)}.
% \end{equation}
where $\left| u_n (\boldsymbol{k}) \right\rangle $ are eigenvectors of the bulk Hamiltonian.
This quantity is well known in one-dimensional systems with inversion symmetry, where it takes two values, zero for the trivial phase and $\pi$ for the topological phase. The most paradigmatic example is the Su-Schrieffer-Heeger model (SSH model) \cite{PhysRevLett.42.1698}.\\

\begin{figure}
    \centering
    \includegraphics[width = 0.35 \textwidth]{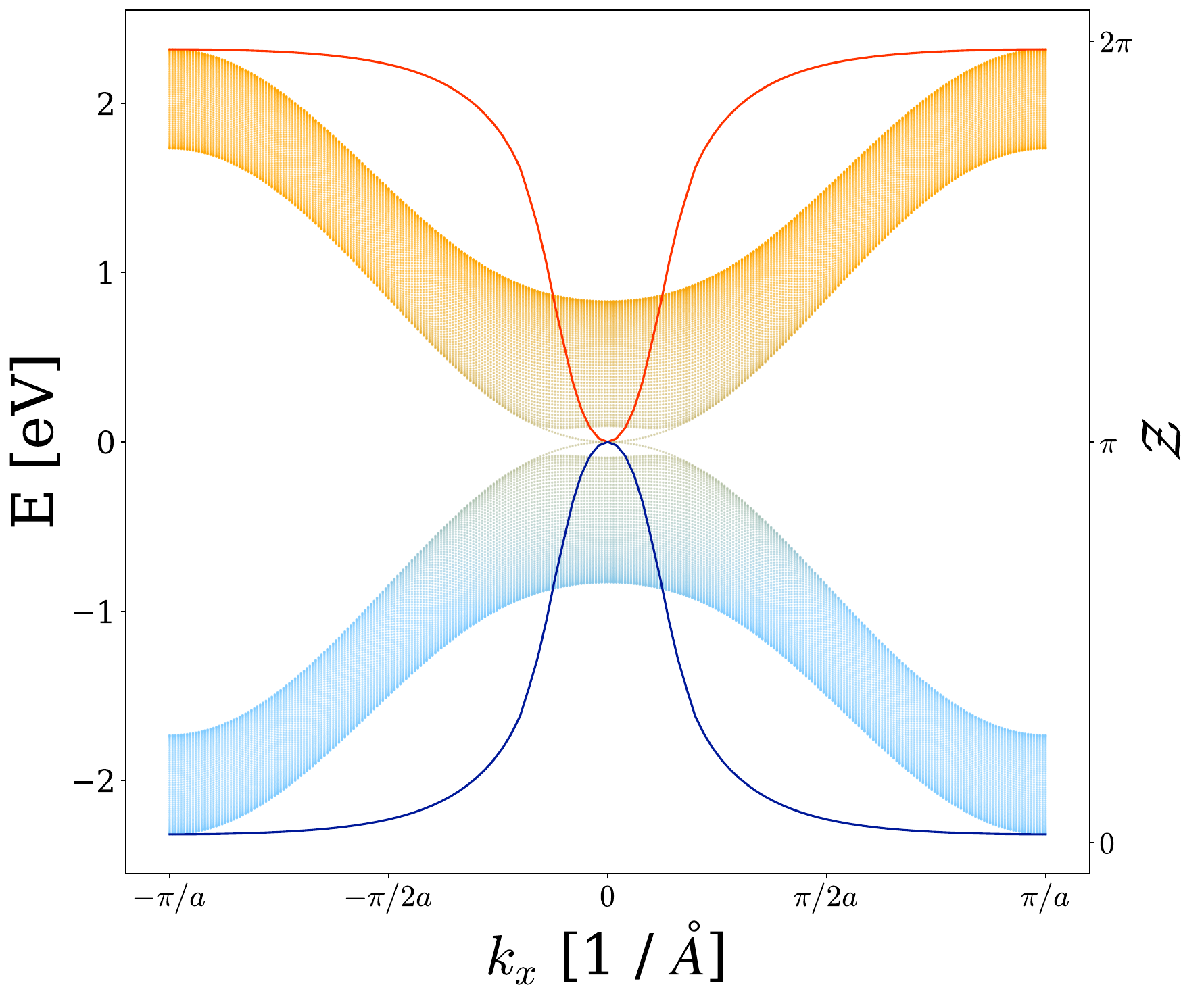}
    \caption{Zak phase (computed in the bulk) of the system in terms of different $k_x$ values overlapped with the band structure of a nanoribbon with finite width in the $y-$direction as a function of $k_x$.}
%    \yuriko{Entiendo que la estructura de bandas es para un nanoribbon, pero la Zak phase en cambio no, verdad? Igual habría que aclarar esto mejor en el texto, citando brevemente la bulk-bouldary correspondence por ejemplo?} \marta{Pendiente de leer el párrafo del texto que va tras la referencia [33,34].}}
    \label{fig:zak-phase}
\end{figure}

% This Zak phase essentially characterizes the topological properties of the system along the chosen direction.
Treating our problem as a $k_x$-dependent one-dimensional model and integrate over all possible values of $k_y$, one can compute this topological invariant along each line of constant $k_x$ within the Brillouin zone. The result, obtained with z2pack package \cite{z2packGresch} for a non-trivial set of values, is shown in Fig.  \ref{fig:zak-phase} for the upper and lower band, where we compare it with the shape of the bands for a finite width nanoribbon. Focusing on the upper band, the Zak phase is maximum for $k_x = 0$ where $\mathcal{Z} = \pi$ and then decays continuously. %Since we are dealing with a one-dimensional calculation embedded in a two-dimensional system 
%with lack of inversion symmetry due to the presence and parity of the $d_z$ term. 
%sigma_x
The Zak phase is not quantized outside the $k_x=0$ and $k_x=\pm \pi/a$ points. This lack of quantization reflects that fact that the symmetry protecting the topology at $k_x=0$ is not present for other values of $k_x$  \cite{Delplace2011,Asboth2016,martisabate2021}. In Fig. \ref{fig:zak-phase} we can examine the correlation between the behavior of the Zak phase and the existance of edge states. For the values of $k_x$ where the edge states mix with the bulk bands the Zak phase approaches zero. Only at $k_x=0$ we expect full topological protection of the edge states according to the bulk-edge correspondence. 
%\marta{This reflects the fact that while the system has topological characteristics in one direction, it lacks the full protection expected in a true one-dimensional topological system. In Fig. \ref{fig:zak-phase} we overlapped the Zak phase with the band structure of the system confined in the $y$-direction in the whole Brillouin zone, to examine the correlation between the behavior of the Zak phase with the existence of these edge states in terms of the bulk-edge correspondance. Particularly, for $k_x$ where the edge states mix with bulk bands the Zak phase approaches to zero while for $k_x = 0$ it takes the non-trivial topological value, $\mathcal{Z} = \pi$.}
% https://arxiv.org/pdf/1605.00549.pdf

% \textcolor{red}{This can be interpreted geometrically in terms of the $\boldsymbol{d}$-vector surface plotted in Fig. \ref{fig:param_surface}. This surface is a revested cylinder that grows from $- \pi / a$ to $0$ and closes on itself in the same way from $0$ to $\pi / a$. It is important to note that it does not strictly enclose the origin, which is consistent with a zero Chern number. Nonetheless, the elliptical path corresponding to $k_x = 0$ does, suggesting a gap closure and a possible phase transition.}
% %for that specific states.
% \begin{figure}[h]
%       \includegraphics[width=0.8\linewidth]{param_surface_v2_cropped.pdf}
   
%     \caption{{\color{purple}This picture is the geometrical interpretation but if somebody is not familiar with that, it doesn't say anything, check if it is better to remove it.}
%         \yuriko{Estoy de acuerdo con esto, igual es un poco complicado entender esta figura... Voy a echarle una pensada}
%         Visualization of the d-vector surface in momentum space. It is a elliptical cylinder revested, it extends from $k_x = -\pi$ to $k_x = 0$ and takes the same path from $k_x = 0$ to $k_x = - \pi$.}
%     \label{fig:param_surface}
% \end{figure}
% This is not surprising since our edge states live only in one of the two possible directions of the space. 
% Nakahara pag 121
Indeed, this zero energy state at $k_x=0$ can be mapped into the zero energy state of a SSH model so it is topologically protected by chiral symmetry, represented, in this case, by the $\sigma_x$ operator \cite{Ryu2002}. To illustrate this correspondence, one can check that a discretize version (employing the substitution $k_i \rightarrow \frac{1}{a} \sin (k_i a)$, $k_i^2 \rightarrow \frac{2}{a^2} (1 - \cos(k_i a))$ \cite{Shen2017_Book}) of our continuous semi-Dirac Hamiltonian with $k_x=0$ can be obtained by applying a rotation $R = \exp (i \pi / 4 \sigma_y)$ in the Pauli matrix space around the $\sigma_y$ axis to a generalized SSH Hamiltonian
%, $R = \exp (i \pi / 4 \sigma_y)$  with $ v = \left( \frac{aM_0V_y}{2M_1} - \frac{V_y}{a} \right)$ and $w = \frac{V_y}{a}$. The matrix representation of $R$ is
%$H_{\text{SD}} = R H_{\text{SSH}} R^{\dagger}$ with 
%\begin{equation}
%    R = \frac{\sqrt{2}}{2} \begin{pmatrix}
%        1 & 1 \\ -1 & 1 
%    \end{pmatrix} .
%\end{equation}
To make this connection we consider that the Hamiltonian of the SSH model can be written as
% \begin{equation}
%     H_{\text{SSH}} =  \begin{pmatrix}
%         0 & v + \frac{w}{a} \cos (ka) - i \frac{w}{a} \sin (ka) \\
%         v + \frac{w}{a} \cos (ka) + i \frac{w}{a} \sin (ka) & 0 
%     \end{pmatrix} ,
% \end{equation}
\begin{equation}
    H_{\text{SSH}} = \left[ v + w_1 \cos (ka) \right] \sigma_x + w_2 \sin (ka) \sigma_y,
\end{equation}
with $v=M_0-2M_1/a^2$, $w_1=M_1/a^2$  and $w_2 =V_y/a$. In the standard SSH model $w_1=w_2$, 
%and that implies that there are hoppings only to nearest neighbors. 
but this generalized SSH model keeps the chiral symmetry and the topological protection of the edge states.
%, the hoppings can be to first and third nearest neighbors but they still keep the chiral symmetry and the topological protection of edge states. 
However, when $k_x \ne 0$, additional terms proportional to $\sigma_z$ appear that break chiral symmetry.
% and imply hoppings to second nearest neighbors. 
The Zak phase is, thus, not quantized, and the edge states are not topologically protected by symmetry. However, the closest the value of $\mathcal{Z}$ is to $\pi$, the more localized are the corresponding edge states.
%and carry out a discretization of our continuum model, employing the substitution $k_i \rightarrow \frac{1}{a} \sin (k_i a)$, $k_i^2 \rightarrow \frac{2}{a^2} (1 - \cos(k_i a))$ \cite{Shen2017_Book}.}
%%%% https://arxiv.org/pdf/2107.09146.pdf

% That are v = \left( M_0 - \frac{2M_1}{a^2} \right), w = \frac{V_y}{a} with \frac{2M_1}{a^2} = \frac{V_y}{a}.

% R se traduce en la matriz de rotación \begin{equation}\frac{\sqrt{2}}{2}\begin{pmatrix} 1 & 1 \\ -1 & 1 \end{pmatrix} \end{equation}

%\yuriko{SUGGESTION: creo que el tema del SSH es interesante, entiendo que no vale la pena la larga discusión, pero quizás poner el Hamiltoniano SSH habitual y la parametrización que hay que hacer para pasar este al SSH? (Junto a una matriz de rotación, si no me equivoco)}

%Even though states with $k_x \neq 0$ are not protected, they do not immediately delocalized from the edge in the presence of disorder, since the impact of disorder is less pronounced compared to higher energy states. 

% \begin{align}
%     &H_{\text{ssh}} (\boldsymbol{k}) = - (t - \delta t) \sin (k) \sigma_x
%     + \left[ \delta t + (t - \delta t) (1 - \cos (k)) \right] \sigma_z.\\
%     % & H_{SD} (k_x = 0, k_y) = \frac{V_y}{a} \sin (k_y) \sigma_y + \left[ M_0 - \frac{M_1}{a^2} (1 - \cos (k_y)) \right] \sigma_z 
% \end{align}
% \begin{equation} \nonumber
%     \footnotesize
%     \vec{d}_{\text{ssh}} (\boldsymbol{k}) = \begin{pmatrix}
%         -(t-\delta t) k \\
%         0\\
%         \delta t + \frac{(t - \delta t)}{2} k^2 
%     \end{pmatrix}
% \end{equation}

The situation is then similar to high order topological insulators where gapless topological protected states are localized not in the edges but in corners \cite{Schindler2018}. The difference is that in this model the topological protection affects only singular values in momentum space. We can consider this model as an example of high order momentum topological insulator in the reciprocal space. 

\section{Disorder simulations}

In this section we include disorder in the simulations to check the robustness of the particular topological protected zero mode in the semi-Dirac model and compare it with the other edge states with quadratic dispersion.

%Including disorder offers a perspective between the role of symmetry and the robustness of the edge states. 

For our numerical simulations we employ the finite size system of length $L = 50a$ and $W = 66a$.
%, which is enough to let the edge states decay into the bulk. 
We consider Anderson disorder, which is introduced in the Hamiltonian by adding the following onsite term,
\begin{equation}
    H_{A}  = \sum_i \varepsilon_i c_i^\dagger c_i  .
\end{equation}
In this case, the on-site energies $\varepsilon_i$ %is a random quantity 
are random numbers
uniformly distributed in the range $\left[ -\frac{w}{2}, \frac{w}{2}\right]$ where $w$ represents the amplitude of disorder. 
%This disorder induce a random shift in the on-site energy of each lattice site. 
We present the results of an ensemble of 1500 realizations. 
%To account for the statistical nature of the disorder, the results presented are averaged over 1500 realizations, which provides a consistent representation of the system behavior with minimal fluctuations due to randomness.\\

%**I have to add in this paragraph the relation between chiral symmetry and particle-hole symmetric spectrum in this case. \textit{Now, in the SSH model, chiral symmetry is closely related to particle-hole symmetry. As mentioned previously, chiral symmetry implies that the hamiltonian anticommutes with a specific chiral operator, denoted by $\sigma_x$ in this case. This implies that the eigenvalues of the hamiltonian can be classified into pairs with opposite eigenvalues. In other words, if $\left| \Psi \right\rangle$ is a eigenstate with eigenvalue $E$, then $\sigma_x \left| \Psi \right\rangle$ is also a eigenstate with eigenvalue $-E$.} 

Chiral symmetry implies that the Hamiltonian anticommutes with a specific chiral operator, in our model $\sigma_x$. The eigenvalues of the Hamiltonian can be classified into pairs with opposite energy so in this context is the same as particle-hole symmetry. If $\left| \Psi \right\rangle$ is a eigenstate with eigenvalue $E$, then $\sigma_x \left| \Psi \right\rangle$ is also a eigenstate with eigenvalue $-E$. As chiral symmetry is so central in the topological protection of the zero mode in the semi-Dirac Hamiltonian, we want to analyze the difference in the robustness of the edge states and topological properties between a disorder breaking chiral symmetry and a disorder preserving the chiral symmetry. 

\begin{figure}[t]
    \centering
    \includegraphics[width = 0.47\textwidth]{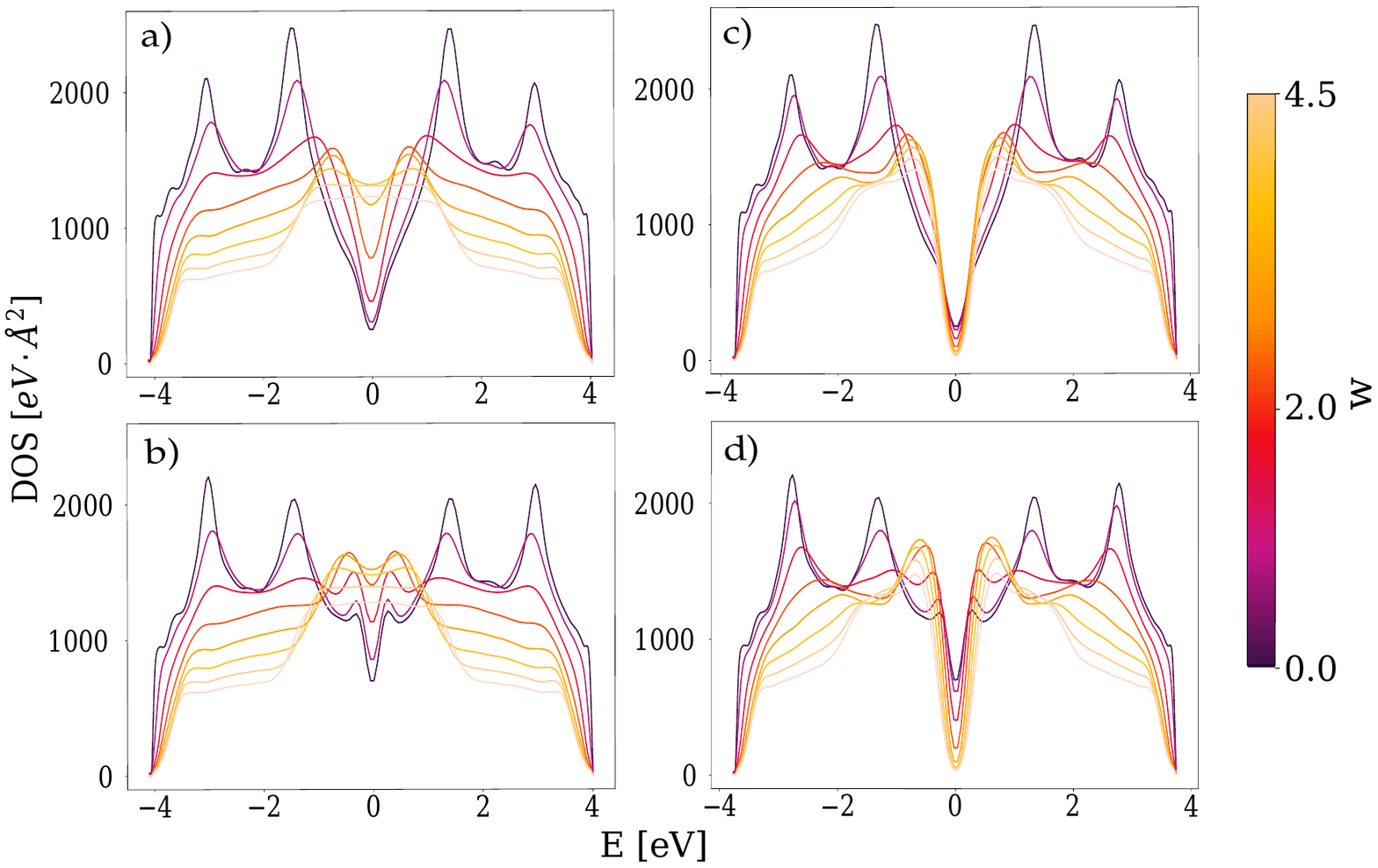}
    \caption{Density of states for different magnitudes of the disorder for the case of disorder breaking particle-symmetry (a), real $\lambda$, (b), complex $\lambda$ and for the case in which the symmetry is preserved, (c) real $\lambda$ and (d) complex $\lambda$.}
    \label{fig:DOS}
\end{figure}

Given that our system has two degrees of freedom per site, an easy way to analyze the impact of losing this symmetry is to implement the disorder in such a way that, in each realization, we can either preserve the particle-hole symmetry by including the Anderson term by
\begin{equation}
    H (\boldsymbol{k}) = H_0 (\boldsymbol{k}) + H_A(w) \sigma_z~,
\end{equation}
or by breaking the symmetry with
\begin{equation}
    H (\boldsymbol{k}) = H_0 (\boldsymbol{k}) + H_A(w) \sigma_0~.
\end{equation}
Among other things, preserving this symmetry ensures that disorder does not cause the overlap of the electron and hole states. In Fig. \ref{fig:DOS} we show the ensemble average density of states for both types of disorder and comparing also results for real and complex values of the decay length $\lambda$. The figure shows the effect close to $E=0$ of the mixing between the particle and hole sectors. When the disorder breaks the particle-hole symmetry, the dip in the DOS at zero energy is lost. Although the qualitative behavior as a function of disorder is similar for real and complex values of $\lambda$, the destruction of the dip is seen more clearly for complex decay lengths. One may notice that in the absence of disorder, there are more states close to zero energy for complex decay lengths than for real decay lengths as the result of parametric dependence. 

%This means that if it is breaking, the mixing of states can lead to a broadening of the energy levels at the Fermi energy as depicted in Fig. \ref{fig:DOS}. 
% \begin{figure}
%         \centering
%         \includegraphics[width=0.4\textwidth]{DOS_rs.pdf}
%         \includegraphics[width=0.4\textwidth]{DOS_rss.pdf}
%         \caption{Density of states in terms of magnitude disorder for the case of disorder that breaks particle-symmetry in each iteration (**top figure) and the model where the particle-hole symmetry is not breaking in any case (bottom figure).}
%     \label{fig:DOS}
% \end{figure}
%As it is shown in Fig. \ref{fig:DOS} the qualitative response of the complex and real cases to disorder is similar, but it is observed that in the absence of disorder the complex case has more states at zero energy than in the real case. Being this states significatively affected by the disorder.\\

\begin{figure*}[ht]
    \fbox{
    \begin{minipage}{0.4104\textwidth}
        \includegraphics[width = \textwidth]{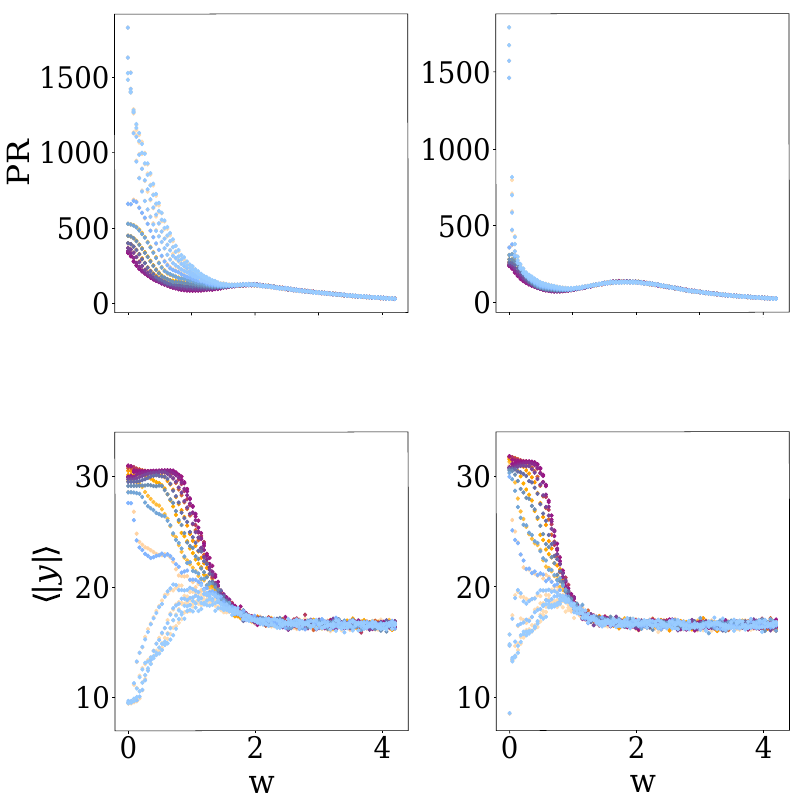}
    \end{minipage}}
    \fbox{
    \begin{minipage}{0.52\textwidth}
        \includegraphics[width = \textwidth]{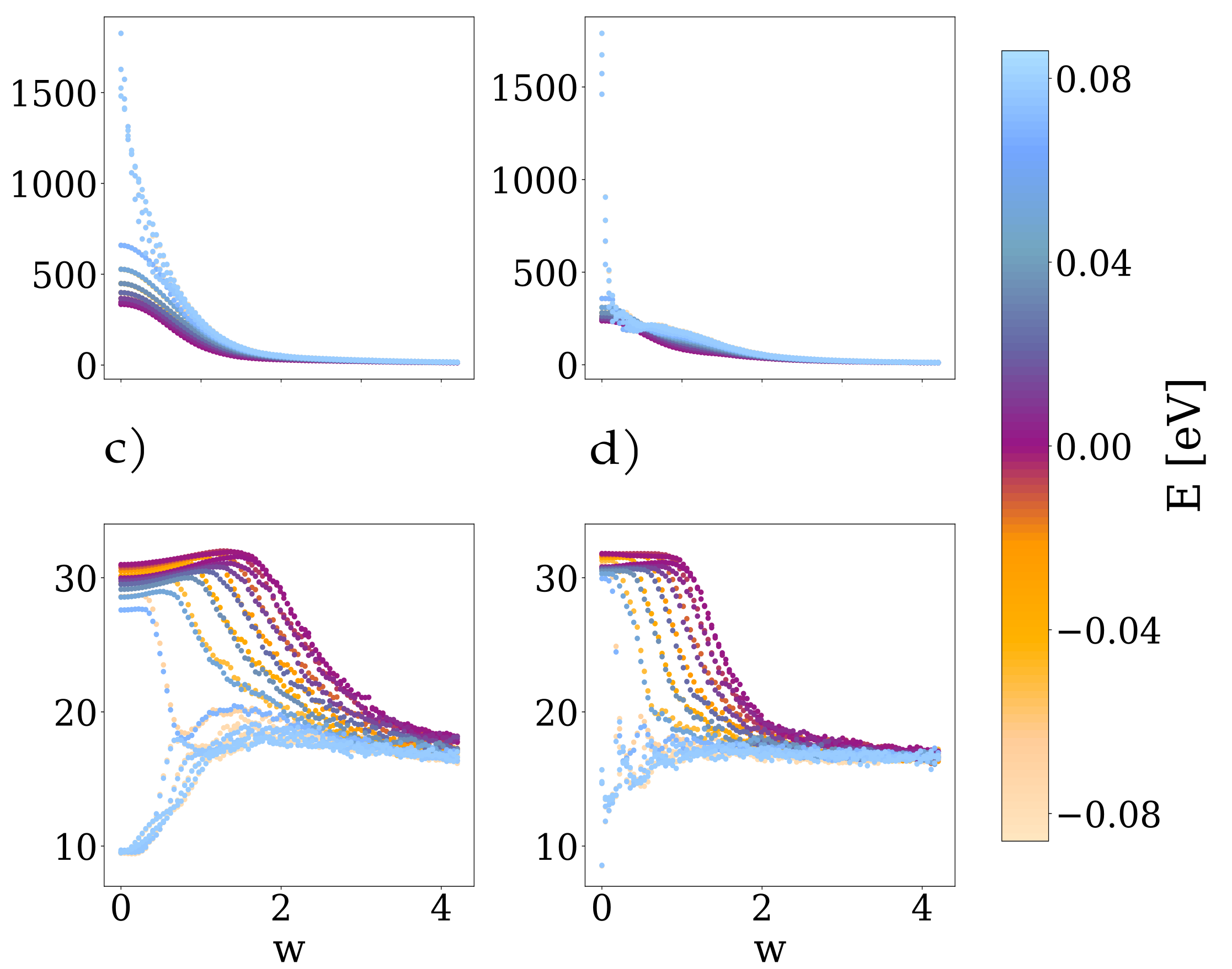}
    \end{minipage}}
    \caption{Participation ratio, PR, and expected value of $y$-position, $\left\langle y \right\rangle$, in terms of the magnitude of disorder for 22 eigenstates around zero energy. The first box contains the results for the case where the disorder realizations break the p-h symmetry and the second inset shows the results when the disorder preserves this symmetry. (a) purely real decay length breaking the symmetry, (b) complex decay length breaking the symmetry, (c) purely real decay length preserving the symmetry, (d) complex decay length preserving the symmetry.}
    \label{fig:yval_pr}
\end{figure*}

To study the impact of disorder on the edge states close to the topologically protected zero mode, we perform a numerical analysis by direct diagonalization of the tight-binding Hamiltonian. The ensemble average expected value of the centroid of the density of the states in the $y$-direction and their participation ratio (PR) are computed as functions of the disorder magnitude. The PR is defined as
\begin{equation}
    \text{PR} = \left( \sum_{i=1}^N p_i\right)^2 \bigg/ \sum_{i=1}^N p_i^2 ,
\end{equation}
where $p_i=\left| \left<i|\Psi \right> \right|^2$ and $i$ is the site index of the tight-binding discretization.  
The PR provides a measure of the localization of the wave functions. A high PR suggest a uniform distribution across all sites, while a low PR indicates localization. We study the case of real and complex nature of the decay length to explore potential differences. However, for the studied value of the width where the coupling between opposite edges is almost absent and the differences are mainly the result of parametric dependence.
We define the centroid of the density as 
\begin{equation}\label{eq:centroid}
\left\langle | y |\right\rangle =\int^{L/2}_{-L/2} \int^{W/2}_{-W/2} |y|~|\Psi(y,x)|^2 dydx.
\end{equation}
%Therefore, the qualitative behavior of the states in both cases is similar,  
The results for both quantities are plotted in Fig. \ref{fig:yval_pr}. The value of $\left\langle y \right\rangle$ distinguishes between edge states centered close to the edge at $y=33a$ and bulk states with $\left\langle y \right\rangle$ close to the center of the nanoribbon. This difference can also be observed in the value of PR. In the edge states there is a transition between edge localization and Anderson localization dominated by disorder effects that corresponds to a transition from $\left\langle y \right\rangle$ close to the edge to $\left\langle y \right\rangle$ inside the bulk. This transition occurs for higher values of the disorder strength $w$ and is much more abrupt for the states closer to $E=0$. This states are also more robust to perturbations due to symmetry-preserving disorder (left block) than to perturbations breaking the chiral symmetry (right block). 

\begin{figure*}[ht]
    \begin{minipage}{0.46\textwidth}
        \centering
        \includegraphics[width = \textwidth]{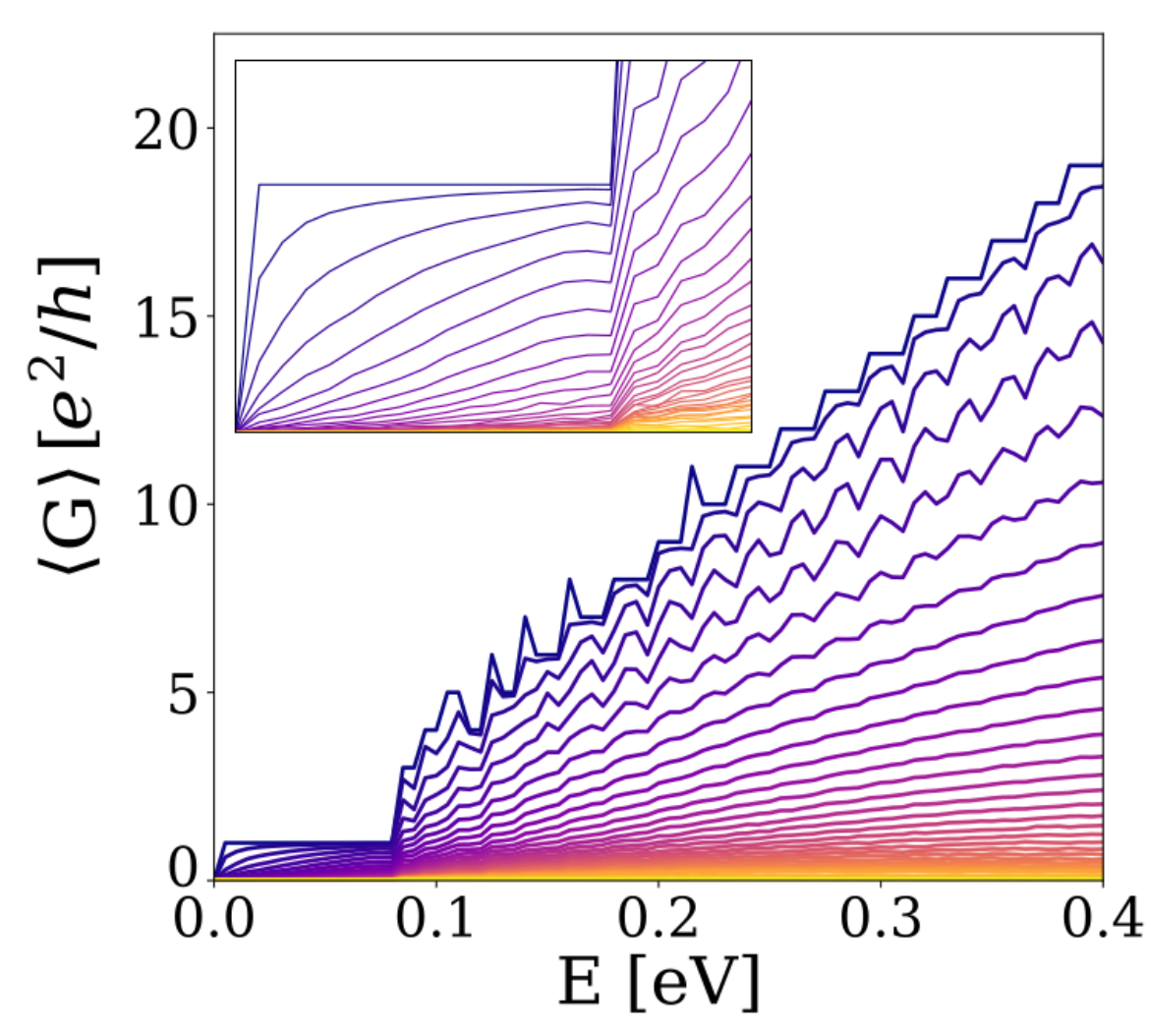}
    \end{minipage}
    \begin{minipage}{0.53\textwidth}
        \centering
        \includegraphics[width = \textwidth]{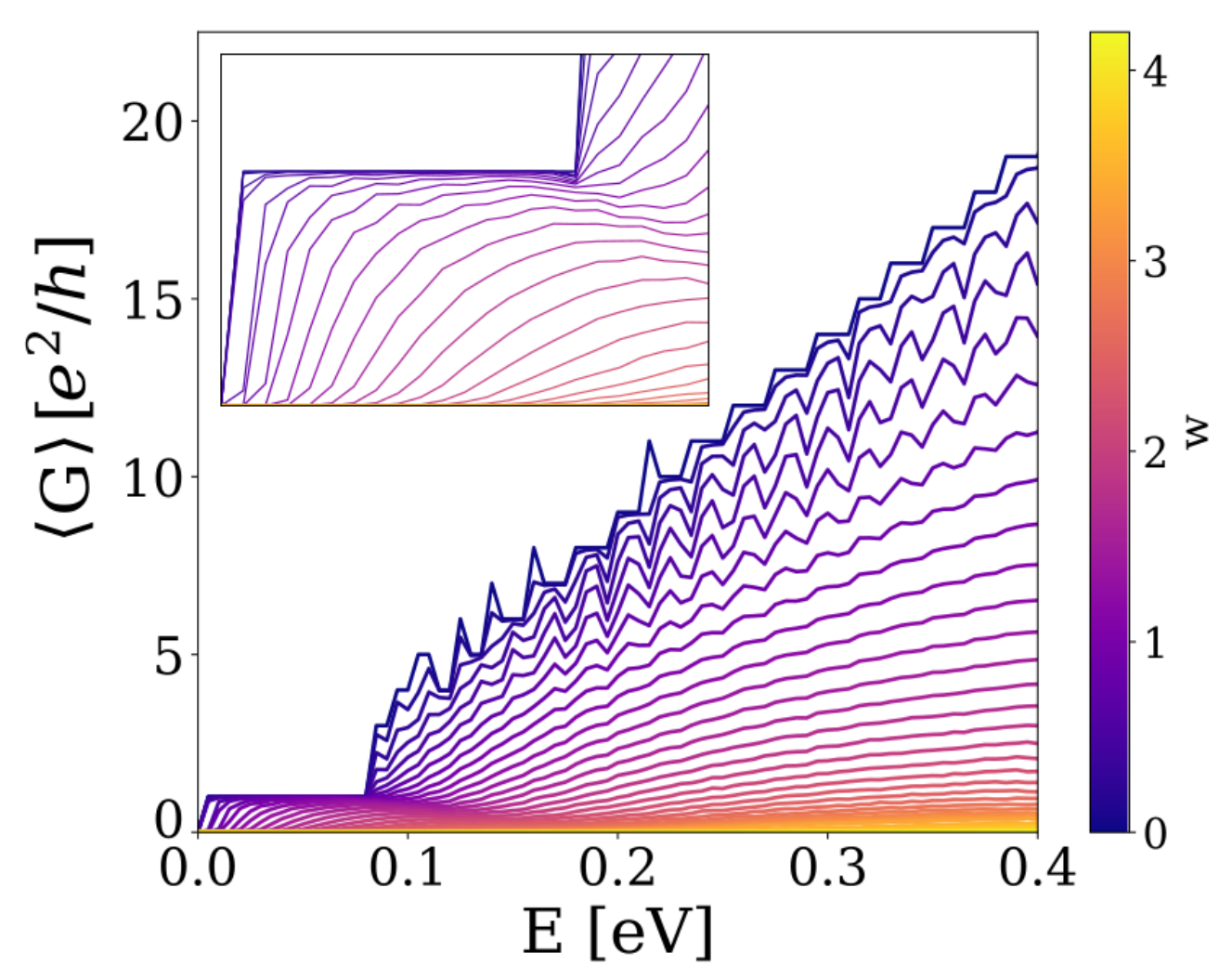}
    \end{minipage}    
        \caption{Conductance averaged over 1500 realizations of disorder in terms of the energy for different magnitude of disorders. On the left is the case in which symmetry is preserved, while on the right is the case in which symmetry is broken.}
        \label{fig:Gy_real}
\end{figure*}

We also show the conductance as a function of disorder in a two-probe setup.  
%in the transport response analyzing the conductance robustness against disorder in a two-probe configuration. 
Conductance is computed using the Landauer-B\"uttiker formalism \cite{Datta1995}, and averaging over disorder realizations. In Fig. \ref{fig:Gy_real} we show the results of the conductance as a function of the Fermi energy of the contacts for the two different types of disorder, symmetry-breaking (left) and symmetry-preserving (right). We only show the results for the parameters given the real decay length, the case with the complex decay length being very similar. The different colors in the figure show the curves as a function of the disorder strength.
%cases considered for a parameter set with real decay length for different magnitudes of disorder until its suppression. value
We can see that conductance steps appear at integer values of $e^2/h$, since we are considering only one spin channel. The robustness of the first quantization step as a function of disorder it is quite remarkable, specially for the symmetry-preserving case.

%\begin{figure}[h]
%    \centering
%    \includegraphics[width = 0.48\textwidth]{G_vs_wcrit_at_E_0.04.pdf}
%    \caption{Averaged conductance in terms of the magnitude of disorder for $E = 0.04~\text{eV}$. The case where the symmetry is preserved, is plotted as a yellow continuous line while when the symmetry is breaking it is represented by a dashed orange line.}
%    \label{fig:G_vs_w}
%\end{figure}
%\yuriko{Creo que estaría bien dar una número de W para el que se pierda la cuantización en un caso y en el otro. Me resulta complicado comparar los valores para el mapa de la figura 8. O quizás diciendo que las curvas están espaciadas para W = 0, 0.1, 0.2 (me lo invento).
%Creo que sería una comparación interesante porque las curvas son muy distintas frente a W pero cuesta cuantificarlo. Otra idea es una figura extra de G(E=0.02) frente a W para los dos casos.}
%Although they are not perfectly quantized, due to finite size effects and folding of the bands, it is not immediately destroyed for weak disorder, the case without symmetry being more affected than the case that preserves symmetry.  

\begin{figure}[ht]
    \centering
    \includegraphics[width = 0.47\textwidth]{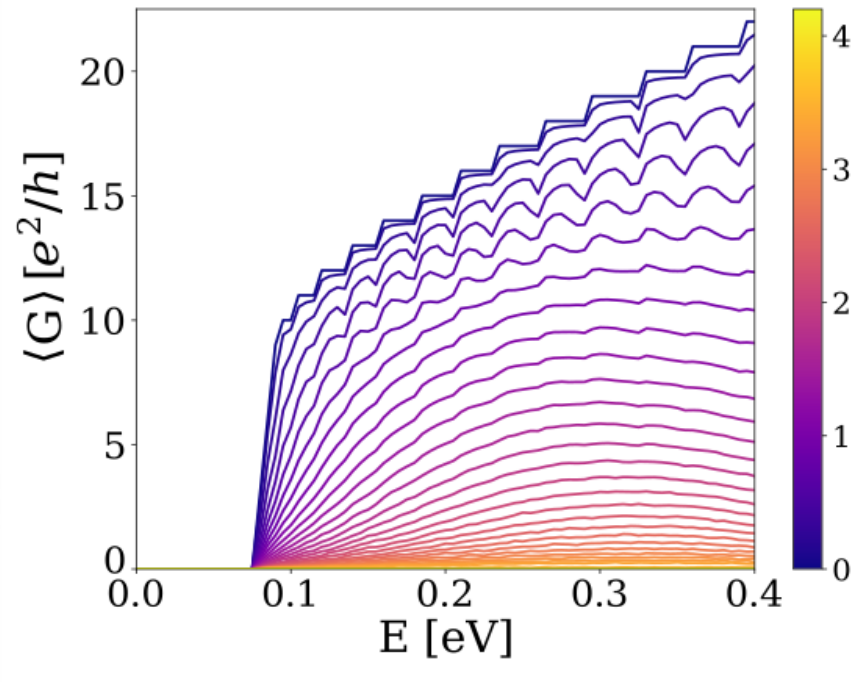}
    \caption{Conductance for a two-probe nanoribbon with leads in $y$-direction and finite width in $x$-direction.}
    \label{fig:Gx_real}
\end{figure}

For comparison, we show in Fig. \ref{fig:Gx_real} the conductance in the $x$-direction with the same parameters and values of disorder. In this case, there are no edge states contributing to the conductance at zero energy and all the transport comes from the bulk.
%**¿Qué pasa con el escalón a 2? electronic rearrangement in contact points.
% \begin{figure}[h]
%         \includegraphics[width=0.2\textwidth]{G_66_real_positive.pdf}
%         \includegraphics[width=0.22\textwidth]{G_vs_w_at_E_-2.35.pdf}
%     \caption{Conductance averaged over 1500 realizations of disorder in terms of the energy for different magnitude of disorders. The case where the symmetry is preserved, is plotted as a continuous line while when the symmetry is breaking it is represented by a dashed line. On the right, the evolution of the averaged conductance in terms of disorder is plotted for an energy in the gap, near bulk and in full band. Shaded line corresponds to the standard deviation of the data}
%     \label{fig:conductance}
% \end{figure}
% The quantized levels that appear in the finite size system are states that an electron can occupy, this means that are going to be closely related with the available channels of the electron transport that contributes to the overall conductance. As a consequence, if the energy levels of complex length, decay in terms of the disorder magnitude before than the real one, means that there is going to be probably less channel transport for the same disorder magnitude so the conductance is going to be less, that is what we see in fig \ref{fig:conductance}. 

\begin{figure}[h]
    \centering
    \includegraphics[width = 0.48\textwidth]{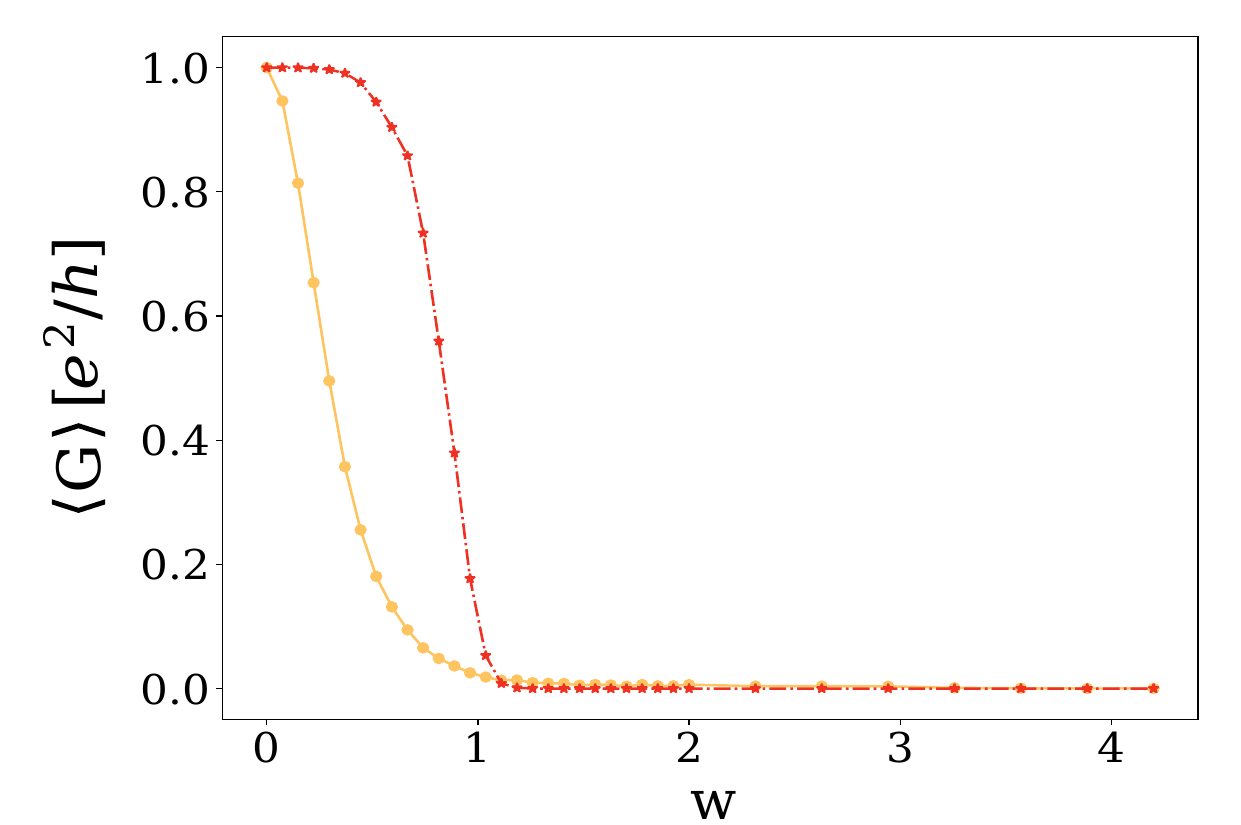}
    \caption{Averaged conductance in terms of the magnitude of disorder for $E = 0.04~\text{eV}$. The case where the symmetry is preserved, is plotted as a dashed orange line while when the case where the symmetry is broken is represented by a continuous yellow line.}
    \label{fig:G_vs_w}
\end{figure}

In Fig. \ref{fig:G_vs_w} we plot the evolution of the averaged conductance for an energy of $0.04~\text{eV}$ (which is at half the first quantization step when $w=0$) as a function of the disorder for the two cases presented. 
%To quantify this difference, 
In order to quantify the sensitivity to disorder of the transport properties of our system we study the value of disorder $w_c$ at which the conductance at $E = 0.04 \text{ eV}$ drops below 60\% of the conductance quantum. In the case without symmetry, this quantity is $w_c = 0.22 \pm 0.04$ while in the case with symmetry, $w_c = 0.81 \pm 0.04$.

\section{Summary and conclusions}

We have studied the anisotropic edge states appearing on a semi-Dirac model using a $k\cdot p$ Hamiltonian that includes a mass and a curvature term. This special dispersion relation is halfway between the typical semiconductor spectrum, with quadratic dependence, and the Dirac spectrum, with linear dependence. As a result, when the parameters of the model present band inversion, edge states are present in the system but only in one direction.

%We have started with an analytical description of the edge states and we combined these result with numerical simulations getting a good fitting between both. 

A simple exponential {\it ansatz} for the description of the edge states enables us to derive their wave functions and dispersion relation. These states exhibit a quadratic dispersion and no chiral properties as a function of the momentum. Their hole or particle character depends on whether they belong to the upper or lower edges of the finite nanoribbon. These properties may suggest that the edge states have no topological character in this model and are consistent with the value of the Chern number $C=0$. We have verified numerically the analytical results. 

We have then explored a different strategy to understand better the topological properties of our anisotropic model and the character of the edge states. Instead of computing the 2D Chern number we have computed a one-dimensional topological number, the Zak phase, for each value of the momentum $k_x$ in the direction of the cuadratic dispersion. We have found a non-trivial Zak phase only for the zero mode state at $k_x=0$. Only these zero modes are protected due to the chiral (or particle-hole) symmetry of the model at this point. This central result of our work is similar to the dimensionality reduction of topological properties that occur in high-order topological insulators \cite{Schindler2018}. However, instead of spatially confined topological protected states like corner states, here we have a state extended in space but localized in momentum.     

From the topological protection exclusive to the $k_x=0$ state, one can infer that there will be a robust quantization step in two-terminal conductance measurements while one does not expect any particular quantum Hall signal due to the zero value of the Chern number. We have presented the localization properties of the edge states and the conductance as a function of the disorder strength for two different kinds of disorder, breaking and preserving the particle-hole symmetry. Our results show that the edge states are quite robust to disorder perturbations, specially if the disorder preserves the particle-hole symmetry. The conductance calculations show indeed the presence of a robust quantization step and the difference between the two types of disorder. 

%The topologically protected states at $k_x = 0$ exhibit a quadratic dependence on momentum around the degeneracy point, as verified analytically and numerically. The non-trivial Zak phase of $\pi$ associated with these states confirms their topological protection due to chiral symmetry.
%However, although the states close to it are not protected, they are localized at the edge and are less robust but robust, and the quantized robust conductance is observed in two and four-probe configurations.
%We examine the interplay between topology and disorder that breaks particle-hole symmetry and that does not break particle-hole symmetry. 
%Our results set that particle-hole symmetry plays a crucial role in the robustness of the edge states which is significantly reduced when it is not present. 

%...

As a final conclusion, our results look deeper into the topological protection of anisotropic 2D materials. Even if the two-dimensional topological number {\it par excellance}, the Chern number, is zero, the topological properties of the edge states may be non-trivial.

\acknowledgments

We acknowledge financial support from
the Ministerio de Ciencia e Innovación of Spain (Spanish Ministry of Science, Innovation, and Universities) and
FEDER (ERDF: European Regional Development Fund) under Research Grants No.
PID2022-136285NB-C31/C32. M.G.O. acknowledges FEDER/Junta de Castilla y
León Research Grant No. SA121P20.

\bibliography{AAbiblioSemiDirac}

\begin{thebibliography}{40}%
\makeatletter
\providecommand \@ifxundefined [1]{%
 \@ifx{#1\undefined}
}%
\providecommand \@ifnum [1]{%
 \ifnum #1\expandafter \@firstoftwo
 \else \expandafter \@secondoftwo
 \fi
}%
\providecommand \@ifx [1]{%
 \ifx #1\expandafter \@firstoftwo
 \else \expandafter \@secondoftwo
 \fi
}%
\providecommand \natexlab [1]{#1}%
\providecommand \enquote  [1]{``#1''}%
\providecommand \bibnamefont  [1]{#1}%
\providecommand \bibfnamefont [1]{#1}%
\providecommand \citenamefont [1]{#1}%
\providecommand \href@noop [0]{\@secondoftwo}%
\providecommand \href [0]{\begingroup \@sanitize@url \@href}%
\providecommand \@href[1]{\@@startlink{#1}\@@href}%
\providecommand \@@href[1]{\endgroup#1\@@endlink}%
\providecommand \@sanitize@url [0]{\catcode `\\12\catcode `\$12\catcode
  `\&12\catcode `\#12\catcode `\^12\catcode `\_12\catcode `\%12\relax}%
\providecommand \@@startlink[1]{}%
\providecommand \@@endlink[0]{}%
\providecommand \url  [0]{\begingroup\@sanitize@url \@url }%
\providecommand \@url [1]{\endgroup\@href {#1}{\urlprefix }}%
\providecommand \urlprefix  [0]{URL }%
\providecommand \Eprint [0]{\href }%
\providecommand \doibase [0]{https://doi.org/}%
\providecommand \selectlanguage [0]{\@gobble}%
\providecommand \bibinfo  [0]{\@secondoftwo}%
\providecommand \bibfield  [0]{\@secondoftwo}%
\providecommand \translation [1]{[#1]}%
\providecommand \BibitemOpen [0]{}%
\providecommand \bibitemStop [0]{}%
\providecommand \bibitemNoStop [0]{.\EOS\space}%
\providecommand \EOS [0]{\spacefactor3000\relax}%
\providecommand \BibitemShut  [1]{\csname bibitem#1\endcsname}%
\let\auto@bib@innerbib\@empty
%</preamble>
\bibitem [{\citenamefont {Novoselov}\ \emph {et~al.}(2004)\citenamefont
  {Novoselov}, \citenamefont {Geim}, \citenamefont {Morozov}, \citenamefont
  {Jiang}, \citenamefont {Zhang}, \citenamefont {Dubonos}, \citenamefont
  {Grigorieva},\ and\ \citenamefont {Firsov}}]{Novoselov2004}%
  \BibitemOpen
  \bibfield  {author} {\bibinfo {author} {\bibfnamefont {K.~S.}\ \bibnamefont
  {Novoselov}}, \bibinfo {author} {\bibfnamefont {A.~K.}\ \bibnamefont {Geim}},
  \bibinfo {author} {\bibfnamefont {S.~V.}\ \bibnamefont {Morozov}}, \bibinfo
  {author} {\bibfnamefont {D.}~\bibnamefont {Jiang}}, \bibinfo {author}
  {\bibfnamefont {Y.}~\bibnamefont {Zhang}}, \bibinfo {author} {\bibfnamefont
  {S.~V.}\ \bibnamefont {Dubonos}}, \bibinfo {author} {\bibfnamefont {I.~V.}\
  \bibnamefont {Grigorieva}},\ and\ \bibinfo {author} {\bibfnamefont {A.~A.}\
  \bibnamefont {Firsov}},\ }\bibfield  {title} {\bibinfo {title} {Electric
  field effect in atomically thin carbon films},\ }\href@noop {} {\bibfield
  {journal} {\bibinfo  {journal} {Science}\ }\textbf {\bibinfo {volume}
  {306}},\ \bibinfo {pages} {666} (\bibinfo {year} {2004})}\BibitemShut
  {NoStop}%
\bibitem [{\citenamefont {Castro~Neto}\ \emph {et~al.}(2009)\citenamefont
  {Castro~Neto}, \citenamefont {Guinea}, \citenamefont {Peres}, \citenamefont
  {Novoselov},\ and\ \citenamefont {Geim}}]{CastroNeto2009}%
  \BibitemOpen
  \bibfield  {author} {\bibinfo {author} {\bibfnamefont {A.~H.}\ \bibnamefont
  {Castro~Neto}}, \bibinfo {author} {\bibfnamefont {F.}~\bibnamefont {Guinea}},
  \bibinfo {author} {\bibfnamefont {N.~M.~R.}\ \bibnamefont {Peres}}, \bibinfo
  {author} {\bibfnamefont {K.~S.}\ \bibnamefont {Novoselov}},\ and\ \bibinfo
  {author} {\bibfnamefont {A.~K.}\ \bibnamefont {Geim}},\ }\bibfield  {title}
  {\bibinfo {title} {The electronic properties of graphene},\ }\href
  {https://doi.org/10.1103/RevModPhys.81.109} {\bibfield  {journal} {\bibinfo
  {journal} {Rev. Mod. Phys.}\ }\textbf {\bibinfo {volume} {81}},\ \bibinfo
  {pages} {109} (\bibinfo {year} {2009})}\BibitemShut {NoStop}%
\bibitem [{\citenamefont {König}\ \emph {et~al.}(2007)\citenamefont {König},
  \citenamefont {Wiedmann}, \citenamefont {Brüne}, \citenamefont {Roth},
  \citenamefont {Buhmann}, \citenamefont {Molenkamp}, \citenamefont {Qi},\ and\
  \citenamefont {Zhang}}]{Koenig2007}%
  \BibitemOpen
  \bibfield  {author} {\bibinfo {author} {\bibfnamefont {M.}~\bibnamefont
  {König}}, \bibinfo {author} {\bibfnamefont {S.}~\bibnamefont {Wiedmann}},
  \bibinfo {author} {\bibfnamefont {C.}~\bibnamefont {Brüne}}, \bibinfo
  {author} {\bibfnamefont {A.}~\bibnamefont {Roth}}, \bibinfo {author}
  {\bibfnamefont {H.}~\bibnamefont {Buhmann}}, \bibinfo {author} {\bibfnamefont
  {L.~W.}\ \bibnamefont {Molenkamp}}, \bibinfo {author} {\bibfnamefont {X.-L.}\
  \bibnamefont {Qi}},\ and\ \bibinfo {author} {\bibfnamefont {S.-C.}\
  \bibnamefont {Zhang}},\ }\bibfield  {title} {\bibinfo {title} {Quantum spin
  hall insulator state in hgte quantum wells},\ }\href@noop {} {\bibfield
  {journal} {\bibinfo  {journal} {Science}\ }\textbf {\bibinfo {volume}
  {318}},\ \bibinfo {pages} {766} (\bibinfo {year} {2007})}\BibitemShut
  {NoStop}%
\bibitem [{\citenamefont {Qi}\ and\ \citenamefont {Zhang}(2011)}]{Qi2011}%
  \BibitemOpen
  \bibfield  {author} {\bibinfo {author} {\bibfnamefont {X.-L.}\ \bibnamefont
  {Qi}}\ and\ \bibinfo {author} {\bibfnamefont {S.-C.}\ \bibnamefont {Zhang}},\
  }\bibfield  {title} {\bibinfo {title} {Topological insulators and
  superconductors},\ }\href@noop {} {\bibfield  {journal} {\bibinfo  {journal}
  {Rev. Mod. Phys.}\ }\textbf {\bibinfo {volume} {83}},\ \bibinfo {pages}
  {1057} (\bibinfo {year} {2011})}\BibitemShut {NoStop}%
\bibitem [{\citenamefont {Maciejko}\ \emph {et~al.}(2011)\citenamefont
  {Maciejko}, \citenamefont {Hughes},\ and\ \citenamefont
  {Zhang}}]{QSHEreview}%
  \BibitemOpen
  \bibfield  {author} {\bibinfo {author} {\bibfnamefont {J.}~\bibnamefont
  {Maciejko}}, \bibinfo {author} {\bibfnamefont {T.~L.}\ \bibnamefont
  {Hughes}},\ and\ \bibinfo {author} {\bibfnamefont {S.-C.}\ \bibnamefont
  {Zhang}},\ }\bibfield  {title} {\bibinfo {title} {The quantum spin hall
  effect},\ }\href@noop {} {\bibfield  {journal} {\bibinfo  {journal} {Annual
  Review of Condensed Matter Physics}\ }\textbf {\bibinfo {volume} {2}},\
  \bibinfo {pages} {31} (\bibinfo {year} {2011})}\BibitemShut {NoStop}%
\bibitem [{\citenamefont {Burkov}\ and\ \citenamefont
  {Balents}(2011)}]{Burkov2011}%
  \BibitemOpen
  \bibfield  {author} {\bibinfo {author} {\bibfnamefont {A.~A.}\ \bibnamefont
  {Burkov}}\ and\ \bibinfo {author} {\bibfnamefont {L.}~\bibnamefont
  {Balents}},\ }\bibfield  {title} {\bibinfo {title} {Weyl semimetal in a
  topological insulator multilayer},\ }\href
  {https://doi.org/10.1103/PhysRevLett.107.127205} {\bibfield  {journal}
  {\bibinfo  {journal} {Phys. Rev. Lett.}\ }\textbf {\bibinfo {volume} {107}},\
  \bibinfo {pages} {127205} (\bibinfo {year} {2011})}\BibitemShut {NoStop}%
\bibitem [{\citenamefont {Liu}\ \emph {et~al.}(2014)\citenamefont {Liu},
  \citenamefont {Zhou}, \citenamefont {Zhang}, \citenamefont {Wang},
  \citenamefont {Weng}, \citenamefont {Prabhakaran}, \citenamefont {Mo},
  \citenamefont {Shen}, \citenamefont {Fang}, \citenamefont {Dai},
  \citenamefont {Hussain},\ and\ \citenamefont {Chen}}]{Liu_2014}%
  \BibitemOpen
  \bibfield  {author} {\bibinfo {author} {\bibfnamefont {Z.~K.}\ \bibnamefont
  {Liu}}, \bibinfo {author} {\bibfnamefont {B.}~\bibnamefont {Zhou}}, \bibinfo
  {author} {\bibfnamefont {Y.}~\bibnamefont {Zhang}}, \bibinfo {author}
  {\bibfnamefont {Z.~J.}\ \bibnamefont {Wang}}, \bibinfo {author}
  {\bibfnamefont {H.~M.}\ \bibnamefont {Weng}}, \bibinfo {author}
  {\bibfnamefont {D.}~\bibnamefont {Prabhakaran}}, \bibinfo {author}
  {\bibfnamefont {S.-K.}\ \bibnamefont {Mo}}, \bibinfo {author} {\bibfnamefont
  {Z.~X.}\ \bibnamefont {Shen}}, \bibinfo {author} {\bibfnamefont
  {Z.}~\bibnamefont {Fang}}, \bibinfo {author} {\bibfnamefont {X.}~\bibnamefont
  {Dai}}, \bibinfo {author} {\bibfnamefont {Z.}~\bibnamefont {Hussain}},\ and\
  \bibinfo {author} {\bibfnamefont {Y.~L.}\ \bibnamefont {Chen}},\ }\bibfield
  {title} {\bibinfo {title} {Discovery of a three-dimensional topological dirac
  semimetal, na 3 bi},\ }\href {https://doi.org/10.1126/science.1245085}
  {\bibfield  {journal} {\bibinfo  {journal} {Science}\ }\textbf {\bibinfo
  {volume} {343}},\ \bibinfo {pages} {864–867} (\bibinfo {year}
  {2014})}\BibitemShut {NoStop}%
\bibitem [{\citenamefont {Naumann}\ \emph {et~al.}(2021)\citenamefont
  {Naumann}, \citenamefont {Arnold}, \citenamefont {Medvecka}, \citenamefont
  {Wu}, \citenamefont {Süss}, \citenamefont {Schmidt}, \citenamefont {Yan},
  \citenamefont {Huber}, \citenamefont {Worch}, \citenamefont {Wilde},
  \citenamefont {Felser}, \citenamefont {Sun},\ and\ \citenamefont
  {Hassinger}}]{Naumann2021}%
  \BibitemOpen
  \bibfield  {author} {\bibinfo {author} {\bibfnamefont {M.}~\bibnamefont
  {Naumann}}, \bibinfo {author} {\bibfnamefont {F.}~\bibnamefont {Arnold}},
  \bibinfo {author} {\bibfnamefont {Z.}~\bibnamefont {Medvecka}}, \bibinfo
  {author} {\bibfnamefont {S.-C.}\ \bibnamefont {Wu}}, \bibinfo {author}
  {\bibfnamefont {V.}~\bibnamefont {Süss}}, \bibinfo {author} {\bibfnamefont
  {M.}~\bibnamefont {Schmidt}}, \bibinfo {author} {\bibfnamefont
  {B.}~\bibnamefont {Yan}}, \bibinfo {author} {\bibfnamefont {N.}~\bibnamefont
  {Huber}}, \bibinfo {author} {\bibfnamefont {L.}~\bibnamefont {Worch}},
  \bibinfo {author} {\bibfnamefont {M.}~\bibnamefont {Wilde}}, \bibinfo
  {author} {\bibfnamefont {C.}~\bibnamefont {Felser}}, \bibinfo {author}
  {\bibfnamefont {Y.}~\bibnamefont {Sun}},\ and\ \bibinfo {author}
  {\bibfnamefont {E.}~\bibnamefont {Hassinger}},\ }\bibfield  {title} {\bibinfo
  {title} {Weyl nodes close to the fermi energy in nbas},\ }\href
  {https://doi.org/10.1002/pssb.202100165} {\bibfield  {journal} {\bibinfo
  {journal} {physica status solidi (b)}\ }\textbf {\bibinfo {volume} {259}},\
  \bibinfo {pages} {2100165} (\bibinfo {year} {2021})}\BibitemShut {NoStop}%
\bibitem [{\citenamefont {Pardo}\ and\ \citenamefont
  {Pickett}(2009)}]{Pardo2009}%
  \BibitemOpen
  \bibfield  {author} {\bibinfo {author} {\bibfnamefont {V.}~\bibnamefont
  {Pardo}}\ and\ \bibinfo {author} {\bibfnamefont {W.~E.}\ \bibnamefont
  {Pickett}},\ }\bibfield  {title} {\bibinfo {title} {Half-metallic
  semi-dirac-point generated by quantum confinement in
  ${\mathrm{tio}}_{2}/{\mathrm{vo}}_{2}$ nanostructures},\ }\href@noop {}
  {\bibfield  {journal} {\bibinfo  {journal} {Phys. Rev. Lett.}\ }\textbf
  {\bibinfo {volume} {102}},\ \bibinfo {pages} {166803} (\bibinfo {year}
  {2009})}\BibitemShut {NoStop}%
\bibitem [{\citenamefont {Katayama}\ \emph {et~al.}(2006)\citenamefont
  {Katayama}, \citenamefont {Kobayashi},\ and\ \citenamefont
  {Suzumura}}]{Katayama2006}%
  \BibitemOpen
  \bibfield  {author} {\bibinfo {author} {\bibfnamefont {S.}~\bibnamefont
  {Katayama}}, \bibinfo {author} {\bibfnamefont {A.}~\bibnamefont
  {Kobayashi}},\ and\ \bibinfo {author} {\bibfnamefont {Y.}~\bibnamefont
  {Suzumura}},\ }\bibfield  {title} {\bibinfo {title} {Electric conductivity of
  the zero-gap semiconducting state in $\alpha$-({BEDT-TTF})$_2${I}$_3$ salt},\
  }\href@noop {} {\bibfield  {journal} {\bibinfo  {journal} {J. Phys. Soc.
  Jap.}\ }\textbf {\bibinfo {volume} {75}},\ \bibinfo {pages} {023708}
  (\bibinfo {year} {2006})}\BibitemShut {NoStop}%
\bibitem [{\citenamefont {Castellanos-Gomez}\ \emph {et~al.}(2014)\citenamefont
  {Castellanos-Gomez}, \citenamefont {Vicarelli}, \citenamefont {Prada},
  \citenamefont {Island}, \citenamefont {Narasimha-Acharya}, \citenamefont
  {Blanter}, \citenamefont {Groenendijk}, \citenamefont {Buscema},
  \citenamefont {Steele}, \citenamefont {Alvarez}, \citenamefont {Zandbergen},
  \citenamefont {Palacios},\ and\ \citenamefont {van~der
  Zant}}]{CastellanosGomez2014}%
  \BibitemOpen
  \bibfield  {author} {\bibinfo {author} {\bibfnamefont {A.}~\bibnamefont
  {Castellanos-Gomez}}, \bibinfo {author} {\bibfnamefont {L.}~\bibnamefont
  {Vicarelli}}, \bibinfo {author} {\bibfnamefont {E.}~\bibnamefont {Prada}},
  \bibinfo {author} {\bibfnamefont {J.~O.}\ \bibnamefont {Island}}, \bibinfo
  {author} {\bibfnamefont {K.~L.}\ \bibnamefont {Narasimha-Acharya}}, \bibinfo
  {author} {\bibfnamefont {S.~I.}\ \bibnamefont {Blanter}}, \bibinfo {author}
  {\bibfnamefont {D.~J.}\ \bibnamefont {Groenendijk}}, \bibinfo {author}
  {\bibfnamefont {M.}~\bibnamefont {Buscema}}, \bibinfo {author} {\bibfnamefont
  {G.~A.}\ \bibnamefont {Steele}}, \bibinfo {author} {\bibfnamefont {J.~V.}\
  \bibnamefont {Alvarez}}, \bibinfo {author} {\bibfnamefont {H.~W.}\
  \bibnamefont {Zandbergen}}, \bibinfo {author} {\bibfnamefont {J.~J.}\
  \bibnamefont {Palacios}},\ and\ \bibinfo {author} {\bibfnamefont {H.~S.~J.}\
  \bibnamefont {van~der Zant}},\ }\bibfield  {title} {\bibinfo {title}
  {Isolation and characterization of few-layer black phosphorus},\ }\href
  {https://doi.org/10.1088/2053-1583/1/2/025001} {\bibfield  {journal}
  {\bibinfo  {journal} {2D Materials}\ }\textbf {\bibinfo {volume} {1}},\
  \bibinfo {pages} {025001} (\bibinfo {year} {2014})}\BibitemShut {NoStop}%
\bibitem [{\citenamefont {Rodin}\ \emph {et~al.}(2014)\citenamefont {Rodin},
  \citenamefont {Carvalho},\ and\ \citenamefont
  {Castro~Neto}}]{black_ph_strain}%
  \BibitemOpen
  \bibfield  {author} {\bibinfo {author} {\bibfnamefont {A.~S.}\ \bibnamefont
  {Rodin}}, \bibinfo {author} {\bibfnamefont {A.}~\bibnamefont {Carvalho}},\
  and\ \bibinfo {author} {\bibfnamefont {A.~H.}\ \bibnamefont {Castro~Neto}},\
  }\bibfield  {title} {\bibinfo {title} {Strain-induced gap modification in
  black phosphorus},\ }\href {https://doi.org/10.1103/PhysRevLett.112.176801}
  {\bibfield  {journal} {\bibinfo  {journal} {Phys. Rev. Lett.}\ }\textbf
  {\bibinfo {volume} {112}},\ \bibinfo {pages} {176801} (\bibinfo {year}
  {2014})}\BibitemShut {NoStop}%
\bibitem [{\citenamefont {Liu}\ and\ \citenamefont {Lei}(2022)}]{Liu2022}%
  \BibitemOpen
  \bibfield  {author} {\bibinfo {author} {\bibfnamefont {G.}~\bibnamefont
  {Liu}}\ and\ \bibinfo {author} {\bibfnamefont {X.}~\bibnamefont {Lei}},\
  }\bibfield  {title} {\bibinfo {title} {{Semi-Dirac and Dirac-node-arc phases
  in a (112) oriented Cd3As2 film}},\ }\href@noop {} {\bibfield  {journal}
  {\bibinfo  {journal} {Journal of Applied Physics}\ }\textbf {\bibinfo
  {volume} {132}},\ \bibinfo {pages} {224304} (\bibinfo {year}
  {2022})}\BibitemShut {NoStop}%
\bibitem [{\citenamefont {Chan}\ \emph {et~al.}(2023)\citenamefont {Chan},
  \citenamefont {Ang},\ and\ \citenamefont {Ang}}]{Xiao_shotnoise}%
  \BibitemOpen
  \bibfield  {author} {\bibinfo {author} {\bibfnamefont {W.~J.}\ \bibnamefont
  {Chan}}, \bibinfo {author} {\bibfnamefont {L.~K.}\ \bibnamefont {Ang}},\ and\
  \bibinfo {author} {\bibfnamefont {Y.~S.}\ \bibnamefont {Ang}},\ }\bibfield
  {title} {\bibinfo {title} {Quantum transport and shot noise in
  two-dimensional semi-dirac system},\ }\href
  {https://api.semanticscholar.org/CorpusID:257405512} {\bibfield  {journal}
  {\bibinfo  {journal} {Applied Physics Letters}\ } (\bibinfo {year}
  {2023})}\BibitemShut {NoStop}%
\bibitem [{\citenamefont {Montambaux}\ \emph
  {et~al.}(2009{\natexlab{a}})\citenamefont {Montambaux}, \citenamefont
  {Pi{\'e}chon}, \citenamefont {Fuchs},\ and\ \citenamefont
  {Goerbig}}]{Montambaux2009b}%
  \BibitemOpen
  \bibfield  {author} {\bibinfo {author} {\bibfnamefont {G.}~\bibnamefont
  {Montambaux}}, \bibinfo {author} {\bibfnamefont {F.}~\bibnamefont
  {Pi{\'e}chon}}, \bibinfo {author} {\bibfnamefont {J.-N.}\ \bibnamefont
  {Fuchs}},\ and\ \bibinfo {author} {\bibfnamefont {M.~O.}\ \bibnamefont
  {Goerbig}},\ }\bibfield  {title} {\bibinfo {title} {A universal hamiltonian
  for motion and merging of dirac points in a two-dimensional crystal},\ }\href
  {https://doi.org/10.1140/epjb/e2009-00383-0} {\bibfield  {journal} {\bibinfo
  {journal} {The European Physical Journal B}\ }\textbf {\bibinfo {volume}
  {72}},\ \bibinfo {pages} {509} (\bibinfo {year}
  {2009}{\natexlab{a}})}\BibitemShut {NoStop}%
\bibitem [{\citenamefont {Banerjee}\ \emph {et~al.}(2009)\citenamefont
  {Banerjee}, \citenamefont {Singh}, \citenamefont {Pardo},\ and\ \citenamefont
  {Pickett}}]{Banerjee2009}%
  \BibitemOpen
  \bibfield  {author} {\bibinfo {author} {\bibfnamefont {S.~S.}\ \bibnamefont
  {Banerjee}}, \bibinfo {author} {\bibfnamefont {R.~R.~P.}\ \bibnamefont
  {Singh}}, \bibinfo {author} {\bibfnamefont {V.}~\bibnamefont {Pardo}},\ and\
  \bibinfo {author} {\bibfnamefont {W.~E.}\ \bibnamefont {Pickett}},\
  }\bibfield  {title} {\bibinfo {title} {Tight-binding modeling and low-energy
  behavior of the semi-dirac point.},\ }\href
  {https://api.semanticscholar.org/CorpusID:5714758} {\bibfield  {journal}
  {\bibinfo  {journal} {Physical review letters}\ }\textbf {\bibinfo {volume}
  {103 1}},\ \bibinfo {pages} {016402} (\bibinfo {year} {2009})}\BibitemShut
  {NoStop}%
\bibitem [{\citenamefont {Montambaux}\ \emph
  {et~al.}(2009{\natexlab{b}})\citenamefont {Montambaux}, \citenamefont
  {Pi\'echon}, \citenamefont {Fuchs},\ and\ \citenamefont
  {Goerbig}}]{Montambaux2009}%
  \BibitemOpen
  \bibfield  {author} {\bibinfo {author} {\bibfnamefont {G.}~\bibnamefont
  {Montambaux}}, \bibinfo {author} {\bibfnamefont {F.}~\bibnamefont
  {Pi\'echon}}, \bibinfo {author} {\bibfnamefont {J.-N.}\ \bibnamefont
  {Fuchs}},\ and\ \bibinfo {author} {\bibfnamefont {M.~O.}\ \bibnamefont
  {Goerbig}},\ }\bibfield  {title} {\bibinfo {title} {Merging of dirac points
  in a two-dimensional crystal},\ }\href
  {https://doi.org/10.1103/PhysRevB.80.153412} {\bibfield  {journal} {\bibinfo
  {journal} {Phys. Rev. B}\ }\textbf {\bibinfo {volume} {80}},\ \bibinfo
  {pages} {153412} (\bibinfo {year} {2009}{\natexlab{b}})}\BibitemShut
  {NoStop}%
\bibitem [{\citenamefont {Carbotte}\ \emph {et~al.}(2019)\citenamefont
  {Carbotte}, \citenamefont {Bryenton},\ and\ \citenamefont
  {Nicol}}]{Carbotte2019}%
  \BibitemOpen
  \bibfield  {author} {\bibinfo {author} {\bibfnamefont {J.~P.}\ \bibnamefont
  {Carbotte}}, \bibinfo {author} {\bibfnamefont {K.~R.}\ \bibnamefont
  {Bryenton}},\ and\ \bibinfo {author} {\bibfnamefont {E.~J.}\ \bibnamefont
  {Nicol}},\ }\bibfield  {title} {\bibinfo {title} {Optical properties of a
  semi-dirac material},\ }\href
  {https://api.semanticscholar.org/CorpusID:119341791} {\bibfield  {journal}
  {\bibinfo  {journal} {Physical Review B}\ } (\bibinfo {year}
  {2019})}\BibitemShut {NoStop}%
\bibitem [{\citenamefont {Huang}\ and\ \citenamefont
  {Shen}(2023)}]{Huang2023TheGA}%
  \BibitemOpen
  \bibfield  {author} {\bibinfo {author} {\bibfnamefont {Y.}~\bibnamefont
  {Huang}}\ and\ \bibinfo {author} {\bibfnamefont {R.}~\bibnamefont {Shen}},\
  }\bibfield  {title} {\bibinfo {title} {The generation and detection of the
  spin-valley-polarization in semi-dirac materials},\ }\href
  {https://api.semanticscholar.org/CorpusID:264935196} {\bibfield  {journal}
  {\bibinfo  {journal} {Physica Scripta}\ } (\bibinfo {year}
  {2023})}\BibitemShut {NoStop}%
\bibitem [{\citenamefont {Zhu}\ \emph {et~al.}(2012)\citenamefont {Zhu},
  \citenamefont {Cheng},\ and\ \citenamefont {Schwingenschl\"ogl}}]{Zhu2012}%
  \BibitemOpen
  \bibfield  {author} {\bibinfo {author} {\bibfnamefont {Z.}~\bibnamefont
  {Zhu}}, \bibinfo {author} {\bibfnamefont {Y.}~\bibnamefont {Cheng}},\ and\
  \bibinfo {author} {\bibfnamefont {U.}~\bibnamefont {Schwingenschl\"ogl}},\
  }\bibfield  {title} {\bibinfo {title} {Band inversion mechanism in
  topological insulators: A guideline for materials design},\ }\href
  {https://doi.org/10.1103/PhysRevB.85.235401} {\bibfield  {journal} {\bibinfo
  {journal} {Phys. Rev. B}\ }\textbf {\bibinfo {volume} {85}},\ \bibinfo
  {pages} {235401} (\bibinfo {year} {2012})}\BibitemShut {NoStop}%
\bibitem [{\citenamefont {Shen}(2012)}]{Shen2017_Book}%
  \BibitemOpen
  \bibfield  {author} {\bibinfo {author} {\bibfnamefont {S.-Q.}\ \bibnamefont
  {Shen}},\ }\href {https://doi.org/10.1007/978-981-10-4606-3} {\emph {\bibinfo
  {title} {{Topological Insulators}}}},\ \bibinfo {series} {Springer}, Vol.\
  \bibinfo {volume} {187}\ (\bibinfo  {publisher} {Springer},\ \bibinfo {year}
  {2012})\BibitemShut {NoStop}%
\bibitem [{\citenamefont {Thouless}\ \emph {et~al.}(1982)\citenamefont
  {Thouless}, \citenamefont {Kohmoto}, \citenamefont {Nightingale},\ and\
  \citenamefont {den Nijs}}]{Thouless1982}%
  \BibitemOpen
  \bibfield  {author} {\bibinfo {author} {\bibfnamefont {D.~J.}\ \bibnamefont
  {Thouless}}, \bibinfo {author} {\bibfnamefont {M.}~\bibnamefont {Kohmoto}},
  \bibinfo {author} {\bibfnamefont {M.~P.}\ \bibnamefont {Nightingale}},\ and\
  \bibinfo {author} {\bibfnamefont {M.}~\bibnamefont {den Nijs}},\ }\bibfield
  {title} {\bibinfo {title} {Quantized hall conductance in a two-dimensional
  periodic potential},\ }\href {https://doi.org/10.1103/PhysRevLett.49.405}
  {\bibfield  {journal} {\bibinfo  {journal} {Phys. Rev. Lett.}\ }\textbf
  {\bibinfo {volume} {49}},\ \bibinfo {pages} {405} (\bibinfo {year}
  {1982})}\BibitemShut {NoStop}%
\bibitem [{\citenamefont {Liu}\ \emph {et~al.}(2016)\citenamefont {Liu},
  \citenamefont {Zhang},\ and\ \citenamefont {Qi}}]{Liu2016}%
  \BibitemOpen
  \bibfield  {author} {\bibinfo {author} {\bibfnamefont {C.-X.}\ \bibnamefont
  {Liu}}, \bibinfo {author} {\bibfnamefont {S.-C.}\ \bibnamefont {Zhang}},\
  and\ \bibinfo {author} {\bibfnamefont {X.-L.}\ \bibnamefont {Qi}},\
  }\bibfield  {title} {\bibinfo {title} {The quantum anomalous hall effect:
  Theory and experiment},\ }\href@noop {} {\bibfield  {journal} {\bibinfo
  {journal} {Annual Review of Condensed Matter Physics}\ }\textbf {\bibinfo
  {volume} {7}},\ \bibinfo {pages} {301} (\bibinfo {year} {2016})}\BibitemShut
  {NoStop}%
\bibitem [{\citenamefont {Huang}\ \emph {et~al.}(2015)\citenamefont {Huang},
  \citenamefont {Liu}, \citenamefont {Zhang}, \citenamefont {Duan},\ and\
  \citenamefont {Vanderbilt}}]{Huang2015}%
  \BibitemOpen
  \bibfield  {author} {\bibinfo {author} {\bibfnamefont {H.}~\bibnamefont
  {Huang}}, \bibinfo {author} {\bibfnamefont {Z.}~\bibnamefont {Liu}}, \bibinfo
  {author} {\bibfnamefont {H.}~\bibnamefont {Zhang}}, \bibinfo {author}
  {\bibfnamefont {W.}~\bibnamefont {Duan}},\ and\ \bibinfo {author}
  {\bibfnamefont {D.}~\bibnamefont {Vanderbilt}},\ }\bibfield  {title}
  {\bibinfo {title} {Emergence of a chern-insulating state from a semi-dirac
  dispersion},\ }\href {https://doi.org/10.1103/PhysRevB.92.161115} {\bibfield
  {journal} {\bibinfo  {journal} {Phys. Rev. B}\ }\textbf {\bibinfo {volume}
  {92}},\ \bibinfo {pages} {161115} (\bibinfo {year} {2015})}\BibitemShut
  {NoStop}%
\bibitem [{\citenamefont {Wunsch}\ \emph {et~al.}(2008)\citenamefont {Wunsch},
  \citenamefont {Guinea},\ and\ \citenamefont {Sols}}]{GuineaSols2008}%
  \BibitemOpen
  \bibfield  {author} {\bibinfo {author} {\bibfnamefont {B.}~\bibnamefont
  {Wunsch}}, \bibinfo {author} {\bibfnamefont {F.}~\bibnamefont {Guinea}},\
  and\ \bibinfo {author} {\bibfnamefont {F.}~\bibnamefont {Sols}},\ }\bibfield
  {title} {\bibinfo {title} {Dirac-point engineering and topological phase
  transitions in honeycomb optical lattices},\ }\href
  {https://doi.org/10.1088/1367-2630/10/10/103027} {\bibfield  {journal}
  {\bibinfo  {journal} {New Journal of Physics}\ }\textbf {\bibinfo {volume}
  {10}} (\bibinfo {year} {2008})}\BibitemShut {NoStop}%
\bibitem [{\citenamefont {Delplace}\ \emph
  {et~al.}(2011{\natexlab{a}})\citenamefont {Delplace}, \citenamefont {Ullmo},\
  and\ \citenamefont {Montambaux}}]{Delplace_2011}%
  \BibitemOpen
  \bibfield  {author} {\bibinfo {author} {\bibfnamefont {P.}~\bibnamefont
  {Delplace}}, \bibinfo {author} {\bibfnamefont {D.}~\bibnamefont {Ullmo}},\
  and\ \bibinfo {author} {\bibfnamefont {G.}~\bibnamefont {Montambaux}},\
  }\bibfield  {title} {\bibinfo {title} {Zak phase and the existence of edge
  states in graphene},\ }\bibfield  {journal} {\bibinfo  {journal} {Physical
  Review B}\ }\textbf {\bibinfo {volume} {84}},\ \href
  {https://doi.org/10.1103/physrevb.84.195452} {10.1103/physrevb.84.195452}
  (\bibinfo {year} {2011}{\natexlab{a}})\BibitemShut {NoStop}%
\bibitem [{\citenamefont {Zak}(1989)}]{ZakPRL}%
  \BibitemOpen
  \bibfield  {author} {\bibinfo {author} {\bibfnamefont {J.}~\bibnamefont
  {Zak}},\ }\bibfield  {title} {\bibinfo {title} {Berry's phase for energy
  bands in solids},\ }\href {https://doi.org/10.1103/PhysRevLett.62.2747}
  {\bibfield  {journal} {\bibinfo  {journal} {Phys. Rev. Lett.}\ }\textbf
  {\bibinfo {volume} {62}},\ \bibinfo {pages} {2747} (\bibinfo {year}
  {1989})}\BibitemShut {NoStop}%
\bibitem [{\citenamefont {Schindler}\ \emph {et~al.}(2018)\citenamefont
  {Schindler}, \citenamefont {Cook}, \citenamefont {Vergniory}, \citenamefont
  {Wang}, \citenamefont {Parkin}, \citenamefont {Bernevig},\ and\ \citenamefont
  {Neupert}}]{Schindler2018}%
  \BibitemOpen
  \bibfield  {author} {\bibinfo {author} {\bibfnamefont {F.}~\bibnamefont
  {Schindler}}, \bibinfo {author} {\bibfnamefont {A.~M.}\ \bibnamefont {Cook}},
  \bibinfo {author} {\bibfnamefont {M.~G.}\ \bibnamefont {Vergniory}}, \bibinfo
  {author} {\bibfnamefont {Z.}~\bibnamefont {Wang}}, \bibinfo {author}
  {\bibfnamefont {S.~S.~P.}\ \bibnamefont {Parkin}}, \bibinfo {author}
  {\bibfnamefont {B.~A.}\ \bibnamefont {Bernevig}},\ and\ \bibinfo {author}
  {\bibfnamefont {T.}~\bibnamefont {Neupert}},\ }\bibfield  {title} {\bibinfo
  {title} {Higher-order topological insulators},\ }\href
  {https://doi.org/10.1126/sciadv.aat0346} {\bibfield  {journal} {\bibinfo
  {journal} {Science Advances}\ }\textbf {\bibinfo {volume} {4}},\ \bibinfo
  {pages} {eaat0346} (\bibinfo {year} {2018})},\ \Eprint
  {https://arxiv.org/abs/https://www.science.org/doi/pdf/10.1126/sciadv.aat0346}
  {https://www.science.org/doi/pdf/10.1126/sciadv.aat0346} \BibitemShut
  {NoStop}%
\bibitem [{\citenamefont {Trifunovic}\ and\ \citenamefont
  {Brouwer}(2019)}]{Trifunovic2019}%
  \BibitemOpen
  \bibfield  {author} {\bibinfo {author} {\bibfnamefont {L.}~\bibnamefont
  {Trifunovic}}\ and\ \bibinfo {author} {\bibfnamefont {P.~W.}\ \bibnamefont
  {Brouwer}},\ }\bibfield  {title} {\bibinfo {title} {Higher-order
  bulk-boundary correspondence for topological crystalline phases},\ }\href
  {https://doi.org/10.1103/PhysRevX.9.011012} {\bibfield  {journal} {\bibinfo
  {journal} {Phys. Rev. X}\ }\textbf {\bibinfo {volume} {9}},\ \bibinfo {pages}
  {011012} (\bibinfo {year} {2019})}\BibitemShut {NoStop}%
\bibitem [{\citenamefont {Groth}\ \emph {et~al.}(2014)\citenamefont {Groth},
  \citenamefont {Wimmer}, \citenamefont {Akhmerov},\ and\ \citenamefont
  {Waintal}}]{Groth_2014}%
  \BibitemOpen
  \bibfield  {author} {\bibinfo {author} {\bibfnamefont {C.~W.}\ \bibnamefont
  {Groth}}, \bibinfo {author} {\bibfnamefont {M.}~\bibnamefont {Wimmer}},
  \bibinfo {author} {\bibfnamefont {A.~R.}\ \bibnamefont {Akhmerov}},\ and\
  \bibinfo {author} {\bibfnamefont {X.}~\bibnamefont {Waintal}},\ }\bibfield
  {title} {\bibinfo {title} {Kwant: a software package for quantum transport},\
  }\href {https://doi.org/10.1088/1367-2630/16/6/063065} {\bibfield  {journal}
  {\bibinfo  {journal} {New Journal of Physics}\ }\textbf {\bibinfo {volume}
  {16}},\ \bibinfo {pages} {063065} (\bibinfo {year} {2014})}\BibitemShut
  {NoStop}%
\bibitem [{\citenamefont {Gonz\'alez}\ and\ \citenamefont
  {Molina}(2017)}]{Gonzalez2017}%
  \BibitemOpen
  \bibfield  {author} {\bibinfo {author} {\bibfnamefont {J.}~\bibnamefont
  {Gonz\'alez}}\ and\ \bibinfo {author} {\bibfnamefont {R.~A.}\ \bibnamefont
  {Molina}},\ }\bibfield  {title} {\bibinfo {title} {Topological protection
  from exceptional points in weyl and nodal-line semimetals},\ }\href
  {https://doi.org/10.1103/PhysRevB.96.045437} {\bibfield  {journal} {\bibinfo
  {journal} {Phys. Rev. B}\ }\textbf {\bibinfo {volume} {96}},\ \bibinfo
  {pages} {045437} (\bibinfo {year} {2017})}\BibitemShut {NoStop}%
\bibitem [{\citenamefont {Benito-Mat\'{\i}as}\ and\ \citenamefont
  {Molina}(2019)}]{Benito2019}%
  \BibitemOpen
  \bibfield  {author} {\bibinfo {author} {\bibfnamefont {E.}~\bibnamefont
  {Benito-Mat\'{\i}as}}\ and\ \bibinfo {author} {\bibfnamefont {R.~A.}\
  \bibnamefont {Molina}},\ }\bibfield  {title} {\bibinfo {title} {Surface
  states in topological semimetal slab geometries},\ }\href
  {https://doi.org/10.1103/PhysRevB.99.075304} {\bibfield  {journal} {\bibinfo
  {journal} {Phys. Rev. B}\ }\textbf {\bibinfo {volume} {99}},\ \bibinfo
  {pages} {075304} (\bibinfo {year} {2019})}\BibitemShut {NoStop}%
\bibitem [{\citenamefont {Su}\ \emph {et~al.}(1979)\citenamefont {Su},
  \citenamefont {Schrieffer},\ and\ \citenamefont
  {Heeger}}]{PhysRevLett.42.1698}%
  \BibitemOpen
  \bibfield  {author} {\bibinfo {author} {\bibfnamefont {W.~P.}\ \bibnamefont
  {Su}}, \bibinfo {author} {\bibfnamefont {J.~R.}\ \bibnamefont {Schrieffer}},\
  and\ \bibinfo {author} {\bibfnamefont {A.~J.}\ \bibnamefont {Heeger}},\
  }\bibfield  {title} {\bibinfo {title} {Solitons in polyacetylene},\ }\href
  {https://doi.org/10.1103/PhysRevLett.42.1698} {\bibfield  {journal} {\bibinfo
   {journal} {Phys. Rev. Lett.}\ }\textbf {\bibinfo {volume} {42}},\ \bibinfo
  {pages} {1698} (\bibinfo {year} {1979})}\BibitemShut {NoStop}%
\bibitem [{\citenamefont {Gresch}\ \emph {et~al.}(2017)\citenamefont {Gresch},
  \citenamefont {Aut\`es}, \citenamefont {Yazyev}, \citenamefont {Troyer},
  \citenamefont {Vanderbilt}, \citenamefont {Bernevig},\ and\ \citenamefont
  {Soluyanov}}]{z2packGresch}%
  \BibitemOpen
  \bibfield  {author} {\bibinfo {author} {\bibfnamefont {D.}~\bibnamefont
  {Gresch}}, \bibinfo {author} {\bibfnamefont {G.}~\bibnamefont {Aut\`es}},
  \bibinfo {author} {\bibfnamefont {O.~V.}\ \bibnamefont {Yazyev}}, \bibinfo
  {author} {\bibfnamefont {M.}~\bibnamefont {Troyer}}, \bibinfo {author}
  {\bibfnamefont {D.}~\bibnamefont {Vanderbilt}}, \bibinfo {author}
  {\bibfnamefont {B.~A.}\ \bibnamefont {Bernevig}},\ and\ \bibinfo {author}
  {\bibfnamefont {A.~A.}\ \bibnamefont {Soluyanov}},\ }\bibfield  {title}
  {\bibinfo {title} {Z2pack: Numerical implementation of hybrid wannier centers
  for identifying topological materials},\ }\href
  {https://doi.org/10.1103/PhysRevB.95.075146} {\bibfield  {journal} {\bibinfo
  {journal} {Phys. Rev. B}\ }\textbf {\bibinfo {volume} {95}},\ \bibinfo
  {pages} {075146} (\bibinfo {year} {2017})}\BibitemShut {NoStop}%
\bibitem [{\citenamefont {Delplace}\ \emph
  {et~al.}(2011{\natexlab{b}})\citenamefont {Delplace}, \citenamefont {Ullmo},\
  and\ \citenamefont {Montambaux}}]{Delplace2011}%
  \BibitemOpen
  \bibfield  {author} {\bibinfo {author} {\bibfnamefont {P.}~\bibnamefont
  {Delplace}}, \bibinfo {author} {\bibfnamefont {D.}~\bibnamefont {Ullmo}},\
  and\ \bibinfo {author} {\bibfnamefont {G.}~\bibnamefont {Montambaux}},\
  }\bibfield  {title} {\bibinfo {title} {Zak phase and the existence of edge
  states in graphene},\ }\href {https://doi.org/10.1103/PhysRevB.84.195452}
  {\bibfield  {journal} {\bibinfo  {journal} {Phys. Rev. B}\ }\textbf {\bibinfo
  {volume} {84}},\ \bibinfo {pages} {195452} (\bibinfo {year}
  {2011}{\natexlab{b}})}\BibitemShut {NoStop}%
\bibitem [{\citenamefont {Asbóth}\ \emph {et~al.}(2016)\citenamefont
  {Asbóth}, \citenamefont {Oroszlány},\ and\ \citenamefont
  {Pályi}}]{Asboth2016}%
  \BibitemOpen
  \bibfield  {author} {\bibinfo {author} {\bibfnamefont {J.~K.}\ \bibnamefont
  {Asbóth}}, \bibinfo {author} {\bibfnamefont {L.}~\bibnamefont
  {Oroszlány}},\ and\ \bibinfo {author} {\bibfnamefont {A.}~\bibnamefont
  {Pályi}},\ }\href {https://doi.org/10.1007/978-3-319-25607-8} {\emph
  {\bibinfo {title} {A Short Course on Topological Insulators}}}\ (\bibinfo
  {publisher} {Springer International Publishing},\ \bibinfo {year}
  {2016})\BibitemShut {NoStop}%
\bibitem [{\citenamefont {Martí-Sabaté}\ and\ \citenamefont
  {Torrent}(2021)}]{martisabate2021}%
  \BibitemOpen
  \bibfield  {author} {\bibinfo {author} {\bibfnamefont {M.}~\bibnamefont
  {Martí-Sabaté}}\ and\ \bibinfo {author} {\bibfnamefont {D.}~\bibnamefont
  {Torrent}},\ }\href@noop {} {\bibinfo {title} {Zak's phase in non-symmetric
  one-dimensional crystals}} (\bibinfo {year} {2021}),\ \Eprint
  {https://arxiv.org/abs/2107.10144} {arXiv:2107.10144 [cond-mat.mtrl-sci]}
  \BibitemShut {NoStop}%
\bibitem [{\citenamefont {Ryu}\ and\ \citenamefont {Hatsugai}(2002)}]{Ryu2002}%
  \BibitemOpen
  \bibfield  {author} {\bibinfo {author} {\bibfnamefont {S.}~\bibnamefont
  {Ryu}}\ and\ \bibinfo {author} {\bibfnamefont {Y.}~\bibnamefont {Hatsugai}},\
  }\bibfield  {title} {\bibinfo {title} {Topological origin of zero-energy edge
  states in particle-hole symmetric systems},\ }\bibfield  {journal} {\bibinfo
  {journal} {Physical Review Letters}\ }\textbf {\bibinfo {volume} {89}},\
  \href {https://doi.org/10.1103/physrevlett.89.077002}
  {10.1103/physrevlett.89.077002} (\bibinfo {year} {2002})\BibitemShut
  {NoStop}%
\bibitem [{\citenamefont {Datta}(1995)}]{Datta1995}%
  \BibitemOpen
  \bibfield  {author} {\bibinfo {author} {\bibfnamefont {S.}~\bibnamefont
  {Datta}},\ }\bibfield  {title} {\bibinfo {title} {Electronic transport in
  mesoscopic systems}\ }(\bibinfo {year} {1995})\BibitemShut {NoStop}%
\bibitem [{\citenamefont {Zhou}\ \emph {et~al.}(2008)\citenamefont {Zhou},
  \citenamefont {Lu}, \citenamefont {Chu}, \citenamefont {Shen},\ and\
  \citenamefont {Niu}}]{Zhou2008}%
  \BibitemOpen
  \bibfield  {author} {\bibinfo {author} {\bibfnamefont {B.}~\bibnamefont
  {Zhou}}, \bibinfo {author} {\bibfnamefont {H.-Z.}\ \bibnamefont {Lu}},
  \bibinfo {author} {\bibfnamefont {R.-L.}\ \bibnamefont {Chu}}, \bibinfo
  {author} {\bibfnamefont {S.-Q.}\ \bibnamefont {Shen}},\ and\ \bibinfo
  {author} {\bibfnamefont {Q.}~\bibnamefont {Niu}},\ }\bibfield  {title}
  {\bibinfo {title} {Finite size effects of helical edge states in hgte/cdte
  quantum wells}\ }(\bibinfo {year} {2008})\BibitemShut {NoStop}%
\end{thebibliography}%

\appendix

\section{Edge states} \label{ap:edge-states}

This appendix details the calculations involved in analytically deriving the dispersion relation of the edge states \eqref{eq:dispers_relat}. We follow the procedure of Ref. \cite{Zhou2008} considering a nanoribbon finite in $y$-direction and periodic boundary conditions in $x$-direction. In this scenario, $k_x$ is a good quantum number and $k_y$ is replace by Peierls substitution $k_y \rightarrow -i \partial_y$. For simplicity, we consider each edge separately, treating them as isolated edges of the infinite half-planes defined by $y \geq W/2$ for the upper edge and $y \leq -W/2$ for the lower edge. We choose these limits instead of y > 0 and y < 0 by analogy with the numerical implementation of the system. As previously mentioned, we choose wave functions that vanish at the edges, for $y = \pm W/2 $, and decays towards the bulk with decay length $\lambda$. This decay behavior is capture by assuming a spacial dependence of the wave function of the form \text{$\psi^\pm_{k_x}$ \footnotesize $\sim e^{ik_x x} e^{ \mp \lambda ( \pm W/2 - y)} (\alpha~~\beta)^T$}. Introducing this \textit{ansatz} into the Schrödinger equation yields to the following dispersion relation
\begin{equation}
E = \pm \sqrt{\left[M_0 - M_1(k_x^2 - \lambda^2)\right]^2 + V_x^2 k_x^4 - V_y^2 \lambda^2} ,
\end{equation}
where, here, $+$ refers to the upper band and $-$ refers to the lower band, equal for both edges.
By reversing this expression, we obtain four possible solutions for $\pm \lambda_1$ and $\pm \lambda_2$,
\begin{widetext}
\begin{equation} \label{eq:decay_length_i}
    \lambda_{i} = \sqrt{k_x^2 + \frac{V_y^2-2M_0M_1}{2M_1^2} \pm \frac{\sqrt{V_y^4 + 4M_1^2(k_x^2V_y^2-V_y^2M_0/M_1 - V_x^2k_x^4 + E^2)}}{2M_1^2}}, ~~~~ i = 1,2 .
\end{equation}
\end{widetext}
Thus, in order to ensure a proper decay, ($\text{Re} (\lambda) > 0$), the positive values of $\lambda$ describe edge states decaying from the upper edge while negative values of $\lambda$ describe edge states decaying from the lower edge. The solutions has a general form,
\begin{equation}
    \psi^{\pm}_{k_x} \sim A_1 \begin{pmatrix}
        \alpha_1 \\ \beta_1
    \end{pmatrix} e^{\mp \lambda_1(\pm W/2 - y)} + A_2 \begin{pmatrix}
        \alpha_2 \\ \beta_2
    \end{pmatrix} e^{\mp \lambda_2(\pm W/2 - y)} .
\end{equation}
where, here, $+$ in for the upper edge and $-$ for the lower one. The Dirichlet boundary conditions $\psi^{\pm}_{k_x}(y = \pm  W/2) = 0$ provides the following relation 
\begin{equation} \label{eq:APcond_1}
    \alpha_1\beta_2-\beta_2\alpha_1=0
\end{equation}
Then, by introducing this general solution in the Schr\"odinger equation, we can get a relation between these coefficients and the parameters of the model
\begin{eqnarray} \label{eq:APcond_2}
    &\alpha_i^{\pm} = V_x k_x^2 \mp V_y \lambda_i;
    \\
    &\beta_i = {E - M_0 + M_1 (k_x^2 - \lambda_i^2)}
\end{eqnarray}
By combining \eqref{eq:APcond_1} and \eqref{eq:APcond_2} we obtain the dispersion relation of the edge states in terms of the decay lengths,
\begin{equation}
    E^{\pm} = M_0 - M_1 ( k_x^2 - \lambda_1 \lambda_2) \mp M_1\frac{V_x}{V_y}k_x^2 (\lambda_1 + \lambda_2). 
\end{equation}
Now, we perform an expansion around $k_x = 0$, to obtain the final expression
\begin{equation}
    E^{\pm} = M_0 - M_1 (\lambda_1^ {(0)} \lambda_2^ {(0)}) \pm M_1 \frac{V_x}{V_y} (\lambda_1^ {(0)} + \lambda_2^ {(0)}) ,
\end{equation}
where $\lambda_i^{(0)}$ refers to decay lengths evaluate in $k_x = 0$. For the derivatives we have take into account that
\begin{widetext}
    \begin{align}
        & \frac{d \lambda_i}{d k_x} =\frac{1}{\lambda_i} \left. \left[ 2 k_x \mp \frac{8M_1^2 \left( k_x V_y^2 - 2 k_x^3 V_x^2 \right)}{\sqrt{V_y^4 + 4 M_1^2 \left( E^2 - k_x^4 V_x^2 + k_x^2 V_y^2 - V_y^2 M_0 / M_1 \right)}} \right]\right|_{k_x = 0} = 0 \\
        & \lambda_1 \lambda_2 = \sqrt{Q + R}; ~~~~~ \left. \frac{d \lambda_1 \lambda_2}{d k_x} \right|_{k_x = 0} = 0, ~~ \left. \frac{d^2 \lambda_1 \lambda_2}{d k_x^2} \right|_{k_x = 0} = -2 \\
        & ~~~~Q = k_x^4 + k_x^2 \frac{V_y^2 - 2M_0 M_1}{M_1^2} \frac{(V_y^2 - 2M_0 M_1)^2}{4 M_1^4}  \\ \nonumber
        & ~~~~R = \frac{1}{4 M_1^4} \left[ V_y^4 + 4M_1^2  \left( E^2 + k_x^2 V_y^2 - k_x^4 V_x^2 - V_y^2 M_0/ M_1 \right) \right] \nonumber 
    \end{align}
\end{widetext}
\end{document}